\documentclass[11pt,a4paper]{article}
\usepackage{epsfig}
\usepackage{amsmath}
\usepackage{amssymb}
\usepackage{verbatim}
\usepackage{bm}
\usepackage{dsfont}
\usepackage[footnotesize]{caption}
\usepackage{subfigure}
\usepackage{cancel}
\usepackage{cite}
\usepackage{graphicx}
\usepackage{epsfig}
\usepackage{multirow}
\usepackage{tikz}

\renewcommand{\vec}[1]{\boldsymbol{#1}}
\newcommand{\D}{\mathrm{d}}
\newcommand{\E}{\mathrm{e}}

\def\Mp{M_{\text P}}

\newcommand{\TR}{T_\text{rh}}
\newcommand{\RH}{\text{rh}}
\newcommand{\MEV}{\,\text{MeV}}
\newcommand{\GEV}{\,\text{GeV}}

\begin{document}

\begin{minipage}{0.4\textwidth}
 \begin{flushleft}
SISSA 17/2015/FISI\\
TTK-15-11
\end{flushleft}
\end{minipage} \hfill
\begin{minipage}{0.4\textwidth}
\begin{flushright}
\end{flushright}
 \end{minipage}

\vskip 1.5cm

\begin{center}
{\LARGE\bf Constraints on the reheating temperature}
\vskip 0.2cm
{\LARGE\bf from sizable tensor modes}

\vskip 2cm

{\large  Valerie~Domcke$^a$ and Jan Heisig$^b$ }\\[3mm]
{\it{
a SISSA/INFN, 34126 Triest, Italy \\
b RWTH Aachen University, 52056 Aachen, Germany }
}
\end{center}

\vskip 1cm

\begin{abstract}
Despite its importance for modeling the homogeneous hot early universe
very little is experimentally known about the magnitude of the reheating temperature,
leaving an uncertainty of remarkable 18 orders of magnitude.
In this paper we consider a general class of polynomial inflaton potentials up to fourth order.
Employing a Monte Carlo scan and imposing theoretical and experimental constraints we derive 
a robust lower limit on the energy scale at the end of inflation,
$V_\text{end}^{1/4} > 3 \times 10^{15}$~GeV for sizable tensor modes, $r \geq 10^{-3}$.
If the reheating phase is perturbative and matter dominated, this translates into a lower bound 
on the reheating temperature, yielding $\TR > 3 \times 10^8 \; (7 \times 10^2)$~GeV for 
gravitational inflaton decay through a generic dimension five (six) operator. 
\end{abstract}

\newpage

\tableofcontents

\section{Introduction}

The primordial fluctuations of the cosmic microwave background (CMB) provide a unique window to 
energy scales far beyond the energies reached at colliders. With the recent Planck data~\cite{Ade:2015lrj}, 
experiments have reached a remarkable sensitivity in decoding this information. The theoretical interpretation
is necessarily prone to be model-dependent, with much progress in interpreting the recent data in 
terms of numerous specific inflation models, or classes of models, cf.\ Refs.~\cite{Ade:2015lrj, Martin:2014vha, 
Okada:2014lxa} for a non-exhaustive list of examples. In this paper, we attempt to derive constraints on the 
relevant energy scales of the very early universe -- in particular on the energy scale of the vacuum energy 
at the end of inflation and on the reheating temperature -- in a more model-independent way, covering in 
particular all single-field inflationary models which can be approximated by a polynomial of degree four 
and a wide range of perturbative reheating scenarios.

The highest observable energy scale in the history of our universe is given by the potential energy driving 
inflation at the time when the largest scales observable today exited the horizon, 
\begin{equation}
 V_*^{1/4} = \left(\frac{3}{2} \pi^2 A_s \right)^{1/4} r^{1/4} \, \Mp \simeq  \; 3.3 \times 10^{16}~\text{GeV} r^{1/4} \,,
\label{eq_Vstar0}
\end{equation}
where $A_s$ denotes the amplitude of the scalar perturbations generated during inflation and $\Mp$ is the 
reduced Planck mass, $\Mp=2.43\times10^{18}\GEV$. The unknown quantity is the tensor-to-scalar ratio 
$r$ of the primordial fluctuations. Currently, the most stringent bounds on the tensor-to-scalar ratio~$r$ 
are obtained from a joint analysis of the BICEP2, Keck Array and Planck data, yielding $r < 0.12$~\cite{Ade:2015tva}.
Many large-field inflation models predict values of $r$ close to this bound, and upcoming experiments
will probe values of $r$ in the percent region, cf.\ e.g.~\cite{EssingerHileman:2010hh, Errard:2010bn, Sheehy:2011yf}.
In this paper we focus on $r \geq 10^{-3}$, which might be probed in the near future.

During inflation the potential energy decreases as the inflaton field slowly rolls down its scalar potential 
until finally the potential energy density becomes a subdominant component of the total energy density. 
This is the onset of the reheating phase, which produces a thermal bath with some high temperature 
$\TR$ -- the initial condition for the homogeneous hot early universe. The value of the reheating temperature 
is for example decisive for the question if thermal leptogenesis~\cite{Fukugita:1986hr} can be responsible 
for the baryon asymmetry of our universe, which is only possible for 
$T_\RH \gtrsim 10^9$~GeV~\cite{Buchmuller:2004nz}. Moreover, the reheating temperature controls the 
relic abundances of long-lived particles such as gravitinos present in supergravity theories, which, depending 
on their mass, life-time and abundance, have the potential to explain dark matter~\cite{Fayet:1981sq,Pagels:1981ke} 
or to cause serious 
problems~\cite{Weinberg:1982zq,Ellis:1984er,Kawasaki:2004yh,Kawasaki:2004qu,Jedamzik:2006xz,Moroi:1993mb}
for the cosmological evolution of the universe, in particular endangering the success of big bang nucleosynthesis
(BBN).  Depending on the particle spectrum, a number of bounds have been derived on the reheating 
temperature~\cite{Moroi:1993mb,Pradler:2006hh,Asaka:2000zh,Steffen:2008bt,Covi:2010au,Roszkowski:2012nq}.
For example, in the context of the 17-parameter $R$-parity conserving phenomenological minimal supersymmetric
standard model, gravitino dark matter with a stau as next-to-lightest supersymmetric particle is viable for a reheating
temperature of $\TR \lesssim {2} \times 10^9$~GeV~\cite{Heisig:2013sva} {taking into account current collider 
bounds on the particle  spectrum}. However, despite its importance for modeling the very early universe, 
very little is experimentally known about the magnitude of the reheating temperature. It is bounded from 
below by the requirement of successful BBN and from above by the upper bound on the scale of inflation, 
$10~\text{MeV} < \TR < 2 \times 10^{16}~\text{GeV}$, leaving an uncertainty of remarkable 18 orders of magnitude.

With the unprecedented sensitivity of CMB measurements, a number of authors have recently revisited 
constraints on the energy scales of the early universe.
In a data-driven approach, the scalar potential can be reconstructed around the pivot-scale of a given 
experiment~\cite{Lesgourgues:2007gp, Kinney:2002qn}, cf.~\cite{Ade:2015lrj} for a recent analysis. 
This yields rather model-independent information on the scalar potential, however only in the 
observable range, i.e.\ for a few Hubble times or correspondingly for a sub-planckian range of 
the inflaton field. Constraints on the reheating temperature have been derived for given inflation 
models by extracting constraints on the number of e-folds elapsed after the pivot-scale left the 
horizon from CMB measurements~\cite{Martin:2010kz}. Since the total number of e-folds elapsed 
depends (weakly) on the reheating temperature, matching the temperature and Hubble horizon 
today implies bounds on the reheating temperature which can be as stringent as 
$T_\RH \gtrsim 10^8 \,\text{-}\, 10^{10}$~GeV~\cite{Cook:2015vqa}, though highly dependent on 
the underlying inflation model as well as on the equation of state during reheating~\cite{Martin:2014nya}.

In this paper, we pursue a different approach, aiming to reduce in particular the dependence on the 
choice of the inflation model.
Performing a dedicated Monte Carlo 
scan based on polynomial inflaton potentials up to degree four and taking into account the recent 
Planck constraints on the inflationary observables, we  obtain a lower bound on the energy density 
at the end of inflation in the single-field slow-roll approximation, $V_\text{end}^{1/4} > 3 \times 10^{15}$~GeV 
for $r \geq 10^{-3}$ -- remarkably close to the scale of grand unified theories (GUTs).
If the reheating phase is matter dominated, this  translates into a lower bound on the reheating temperature, 
$\TR > 3 \times 10^8 \; (7 \times 10^2)$~GeV for gravitational inflaton decay through a generic dimension 
five (six) operator. Both the bounds on $V_\text{end}$ and $T_\RH$ are strengthened for larger values of $r$.

The remainder of this paper is organized as follows. In Sec.~\ref{sec:2} we present analytical bounds on 
the energy density at the end of inflation and on the reheating temperature. These turn out to be highly 
dependent on the inflation model, motivating a systematic investigation of a broad class of models 
based on a Monte Carlo scan in Sec.~\ref{sec:setup}. The results and discussion of this scan are 
presented in Sec.~\ref{sec:results} before concluding in Sec.~\ref{sec:conclusion}.

\section{Analytic estimates of the energy scales \boldmath{$V_\text{end}$} and $T_\RH$ \label{sec:2}}

Inflation is a well established concept to cure several short-comings of the standard model of 
cosmology, e.g.\ explaining the  homogeneity and flatness of our universe on large 
scales~\cite{Guth:1980zm}. Its predictions have been a striking success story, culminating 
in the measurement of the fluctuations of the CMB by the Planck satellite with remarkable 
precision,  which strongly support the simplest model of single-field slow-roll inflation~\cite{Ade:2015lrj}. 
Here, the dynamics of inflation is governed by a scalar field $\phi$ slowly moving down a scalar 
potential $V(\phi)$, the latter providing a large, nearly constant energy density necessary to 
exponentially inflate the universe,
\begin{equation}
 3 H \dot \phi = - V'(\phi) \,,
\end{equation}
with $H$ denoting the Hubble parameter. Inflation ends at $\phi_\text{end} = \phi(t_\text{end})$ 
when the kinetic energy of the inflaton $\phi$ becomes comparable to the vacuum energy 
$V_\text{end} = V(\phi_\text{end})$. Observations of the CMB contain information about the 
scalar potential $N_*$ Hubble times (e-folds) earlier, when the pivot scale $k_*^{-1}$ of a 
given experiment exited the horizon,
\begin{equation}
 N_* = \int_{t_*}^{t_\text{end}} H \D t \sim 50 \,\text{-}\, 60 \,,
\label{eq_Nstar}
\end{equation}
corresponding to a field value of $\phi_* = \phi(t_*)$ and an energy scale $V_* = V(\phi_*)$.\footnote{More 
precisely, the CMB provides information on the shape of the scalar potential about 10 e-folds around 
this point. \smallskip} 
One of the main open questions in inflationary model building is the nature of the scalar potential between 
$\phi_*$ and $\phi_\text{end}$.

By measuring the properties of the primordial fluctuations in the CMB, in particular the amplitude $A_s$
and the tensor-to-scalar ratio $r$, we can determine (or at least constrain) $V_*$, cf.\ Eq.~\eqref{eq_Vstar0}. 
Can we use this information to constrain the energy scales at the end of inflation and reheating, 
$V_\text{end}$ and $T_\RH$? After the pivot scale exited the horizon, inflation continued for
a finite amount of time and the potential driving it cannot be arbitrary steep, so as not to violate slow-roll 
inflation. This will result in a lower bound on $V_\text{end}$, the value of the scalar potential at the end 
of inflation. This energy density is then converted into the energy density of the initial thermal bath. A 
crucial parameter is the time-scale of this transition, in the simplest setup governed by the decay rate 
of the massive inflaton particle after the end of inflation. Exploiting that the inflaton must at least interact 
gravitationally, the lower bound on $V_\text{end}$ can be thus converted into a lower bound on the 
reheating temperature. In this section, our aim is to analytically examine the lower bound on 
$V_\text{end}(r)$ which will then lead to a lower bound on the reheating temperature, $\TR(r)$.

\subsection{Constraining the energy density at the end of inflation}\label{sec:vendbound}

In this subsection we discuss analytical bounds on the drop of the potential energy
during single-field slow-roll inflation. We discuss the most conservative case as well as 
the case of monomial potentials. These cases serve as reference 
cases for the later more
general study presented in Secs.~\ref{sec:setup} and~\ref{sec:results}.

Let us take two points in the inflationary period 1 and 2, $t_1<t_2$. Using the potential slow-roll 
parameters\footnote{Effects of possible violation of the slow-roll approximation will be
discussed when relevant.}
\begin{equation}
 \epsilon = \frac{\Mp^2}{2} \left( \frac{V' }{V} \right)^2 \,, \quad \eta =  \Mp^2 \frac{V''}{V} \,,
\label{eq:slowrolldef}
\end{equation}
we consider the ratio of the potentials at point 1 and 2,
\begin{equation}
\frac{V_1}{V_2} = \exp\left(\int_{V_2}^{V_1} \frac{\D V}{V}\right) = 
\exp\left(\int_{\phi_2}^{\phi_1}\frac{\sqrt{2\epsilon}}{\Mp}  \D\phi\right)\,.
\label{eq:Vrat}
\end{equation}
In order to obtain a large value for $V_1/V_2$ we require large $\epsilon$ for as
many e-folds, $N$, as possible. On the other hand, in order to allow for large $N$,  $\epsilon$ has to 
stay below $\epsilon_{\text{end}}$ (otherwise inflation ends). Hence, the largest $V_1/V_2$ is 
obtained by a constant $\epsilon$ just below $\epsilon_{\text{end}}$ for as many e-folds
as allowed. For this case the number of e-folds is obtained by
\begin{equation}
N = \int_{\phi_2}^{\phi_1}\frac{1}{\sqrt{2\epsilon}\,\Mp} \D\phi \ge
\frac{\phi_1 -  \phi_2}{ \sqrt{2\epsilon_{\text{end}}}\,\Mp}\,.
\end{equation}
This determines the maximal field excursion allowed by slow-roll inflation.
Hence, we can impose the conservative bound 
\begin{equation}
\frac{V_1}{V_2 } < \E^{2\epsilon_{\text{end}} N }\,.
\label{eq:expbound}
\end{equation}
The corresponding potential yielding the largest $V_1/V_2$ for a fixed number of
e-folds has an exponential shape\footnote{Models with an exponential inflaton potential 
were discussed e.g. in~\cite{Lucchin:1984yf}.}
\begin{equation}
V(\phi) = V_1 \E^{\sqrt{2\epsilon}{(\phi_1-\phi)/\Mp}}\,,
\label{eq:Vexp}
\end{equation}
and fulfills the condition $\eta = 2 \epsilon$.
Note that in this 
case, some other mechanism is needed to end slow-inflation (e.g.\ a second scalar field as 
in hybrid inflation). Requiring $\epsilon, \eta < 1$, Eq.~\eqref{eq:expbound} yields an absolute 
lower bound of $V_*/V_\text{end} < e^{N_*}$, corresponding to a decrease of five orders of 
magnitude in the energy scale for $N_* = 50$. However, as $\epsilon\sim1/2$ is clearly 
excluded for the first 5-10 e-folds constrained by the CMB this bound is conservative and 
the dramatic exponential drop can at most take place for the last $N_*'$ e-folds not 
constrained by the CMB.

Next, let us the consider the situation where $\eta( \phi) = q \, \epsilon ( \phi)$, $q \in \mathbb{R}$. 
This corresponds to the case of a monomial potential $V\propto \phi^p$, $p>0$~\cite{Linde:1983gd}, 
and we can identify $q = 2(p-1)/p$. In this case
\begin{equation}
\frac{V_*}{V_\text{end} } =\left(\frac{\epsilon_\text{end}}{\epsilon_*}\right)^{p/2}
=\left( \frac{4\epsilon_\text{end} N_* + p}{p}\right)^{p/2}\,.
\label{eq:Vratmon}
\end{equation}
Note that Eq.~\eqref{eq:Vratmon} contains the limiting case $|\eta| \ll \epsilon$ for $p\simeq 1$. 
Conversely, if $|\eta| \gg 2 \epsilon$ we obtain $V_*/V_\text{end} \to 1$, since the potential
must be extremely flat.

From this we can draw an important conclusion. In the special case of $\eta = 2 \epsilon = \text{constant}$, 
the amplitude of the scalar potential during the last $N'_* \sim 40 \,\text{-}\, 50$ e-folds can indeed drop dramatically.
However, this relies heavily on the specific exponential shape of the potential and is not the `generic' 
case. Indeed for monomial potentials, Eq.~\eqref{eq:Vratmon} yields  a much more restrictive bound of 
$(V_*/V_\text{end})^{1/4} < 5.2$ for $N_* < 60$ and $p < 3$, whereby $p \geq 3$ is already strongly 
disfavored by the recent Planck data~\cite{Ade:2015lrj}.
In Sec.~\ref{sec:setup} we will hence pursue the ansatz of a fourth order polynomial in order to systematically
study the largest $V_*/V_\text{end}$ that additionally are in agreement with current observations from the
CMB. In this ansatz it is impossible to achieve a constant value of $\eta = 2 \epsilon$ for all values of $ \phi$ 
for non-vanishing $\epsilon$.\footnote{%
To maintain a constant value of $\eta = 2 \epsilon$ and hence saturate the bound 
Eq.~\eqref{eq:expbound} for $N_* \simeq50$ e-folds
a polynomial of order $\sim180$ is needed.}
In Sec.~\ref{sec:results} we will find that within this framework  $V_*/V_\text{end}$
is significantly smaller than the bound~\eqref{eq:expbound}, namely 
$V_*/V_\text{end} \lesssim 10 \;\text{-}\, 1000$, depending on the value of the 
tensor-to-scalar ratio $r$. Moreover we will investigate the fine-tuning required 
in the parameters of the potential in order to saturate this bound.

Before concluding this section, let us add a brief comment on the slow-roll approximation used in this section. 
To obtain large values of $V_*/V_\text{end}$ we are interested in fairly steep potentials.
In particular we will consider potentials which are well described by the slow-roll limit close to $\phi_*$ 
(as dictated by the Planck data) but may have sizable corrections as $\epsilon \sim 1$ is approached 
towards the end of inflation. In other words, the violation of the slow-roll limit is typically accompanied 
by an accelerating inflaton field.
Considering the full equation of motion for the inflaton,
\begin{equation}
 \ddot \phi + 3 H \dot \phi = - V'(\phi)\,,
\label{eq_diff2}
\end{equation}
we note that for monotonic potentials and an accelerating inflaton field, the slow-roll approximation 
over-estimates the velocity $|\dot \phi|$ and hence the distance in field space $| \phi_* - \phi_\text{end}|$. 
This effect is however typically (over)compensated since for an accelerating inflaton field, 
the slow-roll approximation under-estimates the value of $|\phi_* - \phi_\text{end}|$ when 
inflation ends, cf.\ footnote~\ref{ft_slowroll}. In general, the effect is however expected to 
be small since it only concerns the very end of inflation. We will return to this point in Sec.~\ref{sec:Trhbound}.

\subsection{Linking inflation to the reheating phase}\label{sec:linktr}

Inflation is followed by a phase of reheating, during which the vacuum energy is converted into a hot 
thermal bath of standard model particles~\cite{Kofman:1996mv}. In the simplest setup, this implies an 
intermediate stage governed by the coherent oscillations of a heavy scalar field, which then decays 
into the standard model fields with a total decay rate 
$\Gamma_\phi$~\cite{Dolgov:1982th, Albrecht:1982mp ,Abbott:1982hn}. 
This field can but must not be the inflaton field itself.

A scalar field oscillating in a potential $V(\phi) = a \phi^p$ implies an equation of state~\cite{Turner:1983he}
\begin{equation}
 \omega = \frac{\langle p_\phi \rangle}{\langle \rho_\phi \rangle} = \frac{p - 2}{p + 2}\,,
\label{eq:omega}
\end{equation}
where the energy density $\rho_\phi = \dot \phi^2/2 + V(\phi)$ and the pressure 
$p_\phi = \dot \phi^2 - V(\phi)$ are related through the virial theorem, 
$p \, \langle V(\phi) \rangle = 2 \, \langle \dot \phi^2/2 \rangle$. Here we have 
assumed the oscillation frequency $\dot \phi/\phi$ to be much larger than the 
expansion rate $H$, as is the case very soon after the end of inflation. Let us first
assume for simplicity that the end of inflation is followed immediately by a matter 
dominated phase with $\omega = 0$. In other words, we take the potential for the 
scalar field to be essentially quadratic: 
\begin{equation}
 V(\phi) = \frac{1}{2} m_\phi^2 (\phi + c)^2 \quad 
 \text{for } \phi \leq \phi_\text{end} \qquad \text{with }  V(\phi_\text{end}) = V_\text{end}\,,
\label{Vrh}
\end{equation}
thus matching the energy densities at the end of inflation and at the onset 
of reheating.\footnote{Here, we are not specifying the details of the reheating 
mechanism. In particular, this may hold also for scenarios which involve 
tachyonic preheating~\cite{Felder:2000hj}. In this case the large majority of 
the vacuum energy density is rapidly converted into low-momentum modes 
of a symmetry breaking field which again implies an equation of state with 
$\omega = 0$ if the symmetry breaking field is sufficiently heavy.} 
Finally imposing that inflation is indeed over at $\phi_\text{end}$,\footnote{Note that more precisely, 
the end of inflation is reached when the first Hubble slow-roll parameter 
$\epsilon_H =  \dot \phi^2/( 2 H^2)$ reaches $\epsilon_H = 1$~\cite{Liddle:1994dx}. 
As long as slow-roll is a good approximation, $\epsilon_H \simeq \epsilon$. 
At the end of inflation, when slow-roll is violated, $\epsilon_H \geq 1$, one 
finds for an accelerating inflaton field $\epsilon(\phi) > 1$, allowing us the set 
$\epsilon_\text{end} = 1$ in Eq.~\eqref{eq:infend}. In the (untypical) case of an 
decelerating inflaton field at the end of inflation, this only holds approximately. \label{ft_slowroll} }
\begin{equation}
 \epsilon(\phi) \geq \epsilon_\text{end} \qquad \text{for    }  \phi\leq \phi_\text{end}
 \label{eq:infend}
\end{equation}
we find that the mass of the scalar field is bounded from below,
\begin{equation}
\label{eq:inflatonmass}
 m_\phi^2 \geq \epsilon_\text{end} V_\text{end}/\Mp^2 \,.
\end{equation}
If indeed the inflaton field itself is responsible for reheating and the transition from 
inflation to reheating is perturbative, we can match not only the amplitude of the 
potential at $\phi_\text{end}$ but also the first derivative. In this case the bound 
in Eq.~\eqref{eq:inflatonmass} becomes an equality.

For completeness, let us consider also the case that the transition from inflation to 
a matter dominated phase is not instantaneous. For example, picture the more 
general situation that the potential for the scalar field $\phi$ is given by
\begin{equation}
 V(\phi) = \frac{1}{2} m_\phi^2  \phi^2 \, \left(1 + b \frac{|\phi|^{p - 2}}{\Mp^{p-2}}\right) \,,
\end{equation}
for $p > 2$ where for simplicity we have chosen the origin of field-space so as 
$V(0) = 0 = V'(0)$. This corresponds to the situation where just after the end of inflation, 
while the inflaton oscillation is large, the potential is governed by some higher order 
polynomial $\phi^{p}$ until towards the end of reheating, as $\phi \ll \Mp$, the quadratic 
term comes to dominate.
According to Eq.~\eqref{eq:omega}, this models an equation of state with $\omega$ in 
the interval $\{0, 1\}$, in particular including the equation of state of radiation, 
$\omega = 1/3$, corresponding to a quartic potential. We do not consider 
$0 \leq p < 2$ since the linear and constant terms in the scalar potential are 
always absorbed by our definition of the origin of field space. Fractional 
exponents in this range lead to potentials which are not differentiable at their minimum.
 
Matching to $V(\phi_\text{end}) = V_\text{end}$ and $\epsilon(\phi_\text{end}) \geq \epsilon_\text{end}$ 
as above yields
\begin{equation}
 m_\phi^2 \gtrsim \frac{2 V_\text{end}}{b \Mp^2} \left(\frac{2 \epsilon_\text{end}}{p^2} \right)^\frac{p}{2} \,,
\label{eq:mbound2}
\end{equation}
where we have assumed that the $\phi^p$ term dominates in the initial reheating stage 
(else it is irrelevant and Eq.~\eqref{eq:inflatonmass} applies). We immediately see that 
for $b \sim 1$, i.e.\ if the quadratic and $\phi^{p}$ term become comparable at 
$\phi \sim \Mp$ then Eq.~\eqref{eq:mbound2} yields a comparable bound to 
Eq.~\eqref{eq:inflatonmass}. In particular for  $p = 3$ $(p = 4)$ the bound on 
$m_\phi$ is weakened by merely a factor of about 2 (6). If $b \gg 1$, then the bound 
on $m_\phi$ disappears. This corresponds to the limiting case of  a pure $\phi^p$ potential. 
For $b < 1$, Eq.~\eqref{eq:mbound2} seems to improve on the bound of Eq.~\eqref{eq:inflatonmass}, 
note however that in the derivation of Eq.~\eqref{eq:mbound2} we have assumed the 
$\phi^p$ term to be dominant, i.e.\ $b$ is bounded from below. We conclude that as long 
as the quadratic term is not strongly suppressed compared to higher order terms in the 
reheating potential, the bound in Eq.~\eqref{eq:inflatonmass} is not significantly weakened. 
The following results are hence not very sensitive to the assumption of a purely quadratic 
reheating potential, or correspondingly to the assumption of $\omega = 0$. For concreteness, 
we will however focus on $\omega = 0$ in the following.

The reheating temperature is defined as the temperature of the thermal bath when half of 
the total energy density has been converted into radiation. This happens when the decay 
rate of the massive $\phi$-particle becomes comparable to the Hubble rate, 
$H \simeq \Gamma_{\phi}$. Exploiting the expression for the energy density $\rho_\text{r}$ 
of a thermal bath at temperature $T$  as well as the Friedmann equation,
\begin{equation}
 \rho_\text{r} = \frac{\pi^2}{30} g_*  T^4 \,, \quad \rho_{\text{tot}} = 3 H^2 \Mp^2 \,,
\label{trh}
\end{equation}
where $g_*$ denotes the effective number of massless degrees of freedom in the thermal
bath ($g_*^\RH  = 427/4$ for the standard model at high energies), this yields\footnote{Here, 
we use $H = \Gamma_\phi$ and $\rho_r = \rho_\text{tot}/2$  at $T = \TR$. Corrections arise 
due to the relativistic time delay in the decay of the $\phi$-particle and from the precise factor 
$\alpha$ in $\rho_r = \alpha \, \rho_\text{tot}|_{H = \Gamma_\phi}$, which has to be determined 
numerically. However, even taking into account these effects, Eq.~\eqref{trh} typically remains 
a good estimate for the reheating temperature, cf.\ e.g.\ Ref.~\cite{Buchmuller:2013lra}.}
\begin{equation}
\TR = \left( \frac{45}{\pi^2 g_*^\RH} \right)^{1/4} \sqrt{ \Gamma_\phi \Mp}\,.
 \label{eq:Treh}
\end{equation}
A lower bound on this temperature can be obtained by assuming only gravitational interactions 
for the inflaton, resulting in the following effective dimension five operators:
\begin{equation}
\lambda\frac{m_\phi^2}{\Mp}\phi B^* B \;,\quad \lambda \frac{m_\phi}{\Mp}\phi \bar f f \;,
\label{eq:dim5op}
\end{equation}
where $B$ and $f$ are bosonic and fermionic fields, respectively, and $\lambda$ is a
dimensionless coupling.
Such operators may be sourced by a coupling of the inflaton field to the kinetic terms 
of the respective particles.\footnote{%
In the context of supergravity, such effective operators for the decay into a 
supermultiplet $X$ are generated e.g.\ by a term $\lambda\Mp^{-1} \phi |X|^2$
in the K\"ahler potential, which leads e.g.\ to a term 
$\lambda\Mp^{-1} \phi\, \partial_\mu X^* \partial^\mu X$ 
for the corresponding scalar field in the effective Lagrangian, see also~\cite{Endo:2007sz}. 
However, determining the dominant contribution to the decay of the inflaton depends upon 
the SUSY breaking scale and the details of the supersymmetric particle spectrum.}
The decay rate of the non-relativistic particle $\phi$ into a pair of light bosons or fermions is given by
\begin{equation}
 \Gamma_\phi \sim \frac{\lambda^{2} \, m_\phi^3}{16 \pi \Mp^2} \,.
\label{eq:decay}
\end{equation}
In the case of multiple decay modes Eq.~\eqref{eq:decay} turns into the sum over the respective partial
widths.
One might also imagine 
the situation in which dimension-5 operators are forbidden or strongly suppressed. 
In this case there are additional suppression factors of $m_\phi/\Mp$ with respect to 
Eq.~\eqref{eq:dim5op}. 
In summary, this yields
\begin{equation}
\begin{split}
 \TR^{\text{dim}\,5} &\sim \left( \frac{45}{\pi^2 g_*^\RH } \right)^{1/4} \sqrt{\frac{\lambda^2 \, m_\phi^3}{16 \pi \Mp}}\,, \\
 \TR^{\text{dim}\,6} &\sim \left( \frac{45}{\pi^2 g_*^\RH } \right)^{1/4} \sqrt{\frac{\lambda^2 \, m_\phi^5}{16 \pi \Mp^3}}\,.
 \label{eq:TRdim5dim6}
\end{split}
\end{equation}
Plugging Eqs.~\eqref{eq_Vstar0}, \eqref{eq:expbound}, \eqref{eq:inflatonmass} and 
\eqref{eq:TRdim5dim6}, this yields (for the case of the dimension five operator)
\begin{equation}
 \TR^{\text{dim}\,5} \gtrsim 3.7 \MEV \,  
 \lambda \left( \frac{A_s}{2.2 \times 10^{-9}} \right)^{3/4} \left( \frac{r}{0.1} \right)^{3/4} 
\left(\frac{\epsilon_\text{end}}{1/2}\right)^{3/4}
\left(\frac{427/4}{g_*^\RH }\right)^{1/4}\E^{-3/2(N'_* \epsilon_{\text{end}} - 20)   }\,,
\label{eq_Trh_exp}
\end{equation}
for the exponential potential and correspondingly with Eq.~\eqref{eq:Vratmon} for the monomial potentials
\begin{equation}
 \TR^{\text{dim}\,5} \gtrsim 6.0\times 10^{9} \GEV \, 
  \lambda \left( \frac{A_s}{2.2 \times 10^{-9}} \right)^{3/4} \left( \frac{r}{0.1} \right)^{3/4} 
\left(\frac{427/4}{g_*^\RH }\right)^{1/4} \left(\frac{\epsilon_\text{end}}{1}\right)^{3/4} 
\frac{f(r,N_*,\epsilon_{\text{end}})}{0.09} \,,
\label{Trh_mon}
\end{equation}
where $f(r,N_*,\epsilon_{\text{end}}) 
= \left(r/(16 \epsilon_\text{end})\right)^{3 r N_* \epsilon_\text{end}/(32 \epsilon_\text{end} - 2r)}$ 
(for typical values $f(0.1,50,1)\simeq0.09$). Eq.~\eqref{Trh_mon} leads to 
$\TR \gtrsim 10^{9} \GEV$ for $0.001 <  r < 0.1$ {and $\lambda\simeq 1$} with a very mild 
dependence on the choice of $\epsilon_\text{end}$ and $N_*$. On the contrary, in 
Eq.~\eqref{eq_Trh_exp} these parameters enter in the exponent, implying that for 
$N'_* \, \epsilon_\text{end} \gtrsim 20$ the bound does not impose any
restrictions beyond the requirement that thermalization of the universe has to take 
place well before BBN, $\TR>10\MEV$. The huge difference between Eq.~\eqref{eq_Trh_exp} and 
Eq.~\eqref{Trh_mon} stresses the need for a more model-independent approach, which we will 
address in the next section.

Finally, we mention that other mechanisms can prolong the reheating phase by delaying the thermalization 
of the universe due to kinematic effects (such as e.g. kinematic blocking). These effects can further weaken
the bound on $\TR$ but are highly model-dependent \cite{Drewes:2013iaa, Mazumdar:2013gya}.

\section{Bounds for polynomial potentials} \label{sec:setup}

The previous section served to obtain an analytical estimate for a lower bound on the 
reheating temperature. In a next step, we will back this up by performing a systematic 
Monte Carlo scan of suitable inflationary potentials. 
We will specify the class of scalar potentials in Sec.~\ref{subsec:potentials}.
Secs.~\ref{sec_planckdata} and \ref{sec:Nstarcon} are dedicated to the imposed constraints. The procedure 
of the scan will be detailed in Sec.~\ref{scan} while in Sec.~\ref{sec:disk_method} we will discuss 
the properties of the inflationary potentials particularly interesting for exploring the lower bound on the 
reheating temperature. Our results are then presented in Sec.~\ref{sec:results}.

\subsection{Generic inflaton potentials for sizable tensor modes \label{subsec:potentials}}

Our aim in this section is to systematically sample a broad range of inflation models. 
We restrict ourselves to single-field slow-roll inflation models, implying that the potential 
must allow for approximately 50-60 e-folds of inflation and must allow for inflation to end, 
i.e. $\epsilon(\phi_\text{end}) \sim 1$. This in particular excludes situations in which the 
inflaton field becomes trapped in a false vacuum which it can only leave by tunneling 
through a potential barrier or if it is assisted by the dynamics of a further field. Instead, 
we require the potential to be sufficiently `well-behaved', i.e.\ to support a small but strictly 
monotonous slope over a sufficiently large field range. In this spirit, we expand the potential 
around the Planck pivot scale $\phi_*$, where it is well constrained by the CMB data, 
but truncate the expansion at fourth order: 
\begin{equation}
 V(\Delta \phi) \simeq V_* \left(1+ \sum_{n = 1}^4 c_n \, \left(\frac{\Delta \phi}{\Mp}\right)^n\right) 
 \,,\qquad c_n=\frac{\Mp^n}{V_* n!}\frac{\partial^n V}{\partial \phi^n} \bigg|_{\phi_*} \,,
\label{V3}
\end{equation}
with $\Delta \phi = \phi - \phi_*$. The effect of the truncation is of course independent of 
the choice of $\phi_*$, e.g.\ all potentials of the type
\begin{equation}
 V(\phi) = \sum_{n = 0}^4 a_n \phi^n \,
\label{V1}
\end{equation}
are trivially included in the ansatz~\eqref{V3}, including the monomial potentials with integer 
exponent $p \leq 4$. The motivation for this truncation (apart from the obvious effect of practicability) 
is manifold. For small-field inflation models, $\Delta \phi \ll \Mp$, the Taylor expansion of any $V(\phi)$ 
around $\phi_*$ quickly converges and Eq.~\eqref{V3} is a good approximation to the full potential. 
For large-field inflation models, which we are more interested in for the purpose of this paper since 
they come with a high scale of inflation, higher-order operators with sizable coefficients will make 
a sufficiently long period of slow-roll inflation difficult, unless these coefficients are very carefully 
tuned to ensure a cancellation of these contributions. Viable large-field models of inflation hence
typically forbid these terms, e.g.\ by taking the inflaton to be a pseudo Nambu-Goldstone boson as 
in natural inflation~\cite{Freese:1990rb, Adams:1992bn} or, in supersymmetric models, by employing 
a shift-symmetric K\"ahler potential which is only broken by a  renormalizable superpotential~\cite{Kawasaki:2000yn}.
Our ansatz~\eqref{V3} thus covers a large range of known inflation models (cf.\ Ref.~\cite{Ade:2015lrj} for 
a recent overview in light of recent experimental results): Monomial potentials with non-integer power $p$ 
are well approximated by a polynomial of degree four, the crucial quantity $V_*/V_\text{end}$ differs by 
at most a factor of about $0.6 \,\text{-}\, 1.5$ between the pure monomial and its approximation. It moreover 
trivially contains quadratic and quartic hilltop inflation~\cite{Boubekeur:2005zm}, cf.\ Eq.~\eqref{V1}, 
and approximately contains natural inflation, considering $\cos(\phi/f)$ (with $\phi \ll f$) truncated at 
fourth order. In a broader sense, it covers all renormalizable models of inflation. Finally note that typical 
small-$r$ inflation models such hybrid inflation models~\cite{Copeland:1994vg, Linde:1993cn}\footnote{These 
are listed under the name spontaneously broken supersymmetric models in~\cite{Ade:2015lrj}. Even a variant 
of the minimal hybrid inflation model constructed to achieve a large tensor-to-scalar ratio within an intermediate 
field excursion $\Delta \phi_\text{end} \lesssim \Mp$ only yields 
$V_*/V_\text{end} \lesssim 1.05$~\cite{Brummer:2014wxa}.} 
(putting aside the requirement of reaching $\epsilon(\phi_\text{end}) \sim 1$) and Starobinsky-type inflation 
models~\cite{Starobinsky:1980te} (including the original $R^2$ model as well recent rediscoveries of this exponentially
flat potential~\cite{Ellis:2013xoa, Ellis:2013nxa, Buchmuller:2013zfa, Farakos:2013cqa, Ferrara:2013wka}) 
typically feature a very small $V_*/V_\text{end} \lesssim 10$ and are hence not of particular interest for the
lower bound on $V_\text{end}$ and $T_\RH$.

Based on Eq.~\eqref{V3}, we will systematically search for suitable inflation potentials by means of a 
Monte Carlo scan, restricting the coefficients of Eq.~\eqref{V3} by the CMB observations and 
constraining the number of e-folds of inflation $N_*$ by comparing the CMB pivot scale with 
the present Hubble horizon. We take the inflaton to be rolling towards the origin in field-space, 
i.e.\ $\Delta \phi < 0$, and consider the range starting from  $\Delta \phi = 0$ and ending with 
$\epsilon(\Delta \phi_\text{end}) \sim 1$.

\subsection{Constraints from Planck around $\Delta \phi = 0$ \label{sec_planckdata}}

We consider the CMB observables $A_s,n_s,\alpha_s, \kappa_s, r$ which
can be expressed in terms of the potential slow-roll parameters $\epsilon, \eta$
and
\begin{equation}
\xi^2 = \Mp^4  \frac{V'V^{(3)}}{V^2}\,,\quad\sigma^3 = \Mp^6 \frac{{V'}^2V^{(4)}}{V^3}\,,
\label{eq:xisigma}
\end{equation}
to leading order by
\begin{equation}
 \begin{split}
  A_s &= \frac{V}{24 \pi^2 \epsilon \,\Mp^4} \left[1+\mathcal{O}\left(\epsilon\right)\right]\,,\\
   n_s& = 1 - 6 \epsilon + 2 \eta +\mathcal{O}\left(\epsilon^2\right)\,, \\
 \alpha_s &   = 16 \epsilon \eta - 24 \epsilon^2 - 2 \xi^2+\mathcal{O}\left(\epsilon^3\right)\,, \\
 \kappa_s & = -192 \epsilon^3 + 192 \epsilon^2 \eta - 
32 \epsilon \eta^2  - 24 \epsilon \xi^2 + 2 \eta \xi^2 + 2 \sigma^3 +\mathcal{O}\left(\epsilon^4\right)\,,\\
r\, &= 16 \epsilon+\mathcal{O}\left(\epsilon^2\right)\,,
 \end{split}
\label{sr1}
\end{equation}
where the slow-roll parameters are understood to be evaluated at $\Delta \phi = 0$. 
For the computations in our scan we use, however, expressions for the CMB observables 
beyond leading order which are given in the Appendix.
We checked that in our scan the slow-roll parameters
\begin{equation}
\epsilon = \frac{1}{2}c_1^2\,,\quad
\eta  = 2 c_2\,,\quad
\sqrt{\xi^2} = \sqrt{6 c_1 c_3}\,,\quad
\sqrt[3]{\sigma^3} = \sqrt[\leftroot{-1}\uproot{2}\scriptstyle 3]{24 c_1^2 c_4} 
\label{eq:SRcs}
\end{equation}
are indeed sufficiently small compared to one in order to ensure the validity 
of the expansion (see e.g.~\cite{Liddle:1994dx}). This is in particular the case for potentials that allow
for a sufficiently long period of inflation $N_*\gtrsim 50$ and 
values of the potential drop of interest to our work, $V_*/V_\text{end}\gtrsim10$,
where the absolute values of the above parameters are found to be well below 0.1.

The Planck collaboration has provided stringent constraints on these observables, 
which we take into account by calculating the $\chi^2$ of each parameter point 
$\vec v = (\ln A_s, n_s, r, \alpha_s)^T $,
\begin{equation}
 \chi^2(A_s, n_s, r, \alpha_s) = (\vec v - \vec{\bar v})^T \mathcal{C}^{-1} (\vec v - \vec{\bar v}).
\label{eq_chi4}
\end{equation}
Here $\vec{\bar v}$ contains the mean values of the $1 \sigma$ confidence intervals 
measured by Planck,\footnote{Of course the physical meaning of $r$ as the tensor-to-scalar 
ratio implies $ r > 0$ and we will consider only parameter points with $r > 0$.}
\begin{equation}
\begin{split}
  &\ln (10^{10} A_s) = 3.104 \pm 0.035 \,, \quad  n_s = 0.9644 \pm, 0.0049 \,, \\
 &r = 0 \pm 0.069 \,,\quad \alpha_s = -0.0085 \pm 0.0076 \,, \quad \kappa_s = 0.025 \pm 0.013 \,,
\end{split}
\label{eq_planck_obs}
\end{equation}
and $ \mathcal{C}$ 
is the corresponding covariance matrix, $ \mathcal{C} = \vec \sigma^T X \vec \sigma$. Here $\vec \sigma$
is a vector containing the $1 \sigma$ uncertainties of the observables in Eq.\ \eqref{eq_planck_obs} 
and the correlation matrix $X$ encodes the correlations between the observables. Note that here we
assume that $\Delta \phi = 0$ corresponds to the Planck pivot scale, a constraint we will explicitly 
impose in the next section. We use the data 
from the Planck legacy arxiv for the base-model including a non-vanishing running and tensor-to-scalar 
ratio, including the high multipole joint TT, TE and EE constraints and the low multipole polarization 
constraints~\cite{planck_legacy_arxiv}. This corresponds to confidence intervals given in the last 
column of Tab.~4 in the recent Planck analysis~\cite{Ade:2015lrj}.
By reproducing the marginalized joined 68\% and 95\% C.L. constraints in the 
$n_s$-$\alpha_s$- and $n_s$-$r$- plane in \cite{Ade:2015lrj,Planck:2013jfk} 
we checked that the Gaussian approximation
provides a good description of the uncertainties.
At the time of writing, the Planck collaboration had not released an analysis based on a model 
including a non-vanishing $r$, $\alpha_s$ and $\kappa_s$. We thus base our analysis on the 
data provided for the ($r \neq 0$, $\alpha_s \neq 0$, $ \kappa_s = 0$) model. A priori, this could 
be a serious limitation since fairly strong correlations between these parameters have been 
pointed out, e.g.\ the $\alpha_s$-$\kappa_s$ correlation was found to be important in 
\cite{Antusch:2014saa}. However, performing the Monte Carlo scan detailed in Sec.~\ref{scan}, 
we find that the requirement of sustaining inflation for a suitable period (roughly 50 - 65 e-folds, 
see Sec.~\ref{sec:Nstarcon} for details) enforces $-2\times10^{-3}<\kappa_s<10^{-4}$,
independently 
of the much weaker bound in Eq.~\eqref{eq_planck_obs}, cf.\ left panel of Fig.~\ref{fig:contri1}.
This justifies the use of Eq.~\eqref{eq_chi4}, i.e.\ taking into account the correlations between 
$A_s, n_s, \alpha_s$ and $r$ but varying $\kappa_s$ independently.
In the following figures, 
 we show the resulting 95$\%$  confidence region, corresponding to $\chi^2 < 9.72$.

\bigskip

\subsection{Constraints on $N_*$}\label{sec:Nstarcon}

In the previous section we have focused on how the Planck data constrains 
the scalar potential locally, i.e.\ around $\Delta \phi = 0$. Furthermore, the 
scalar potential must fulfill an important global property, it must sustain a 
suitable amount of inflation between $\Delta \phi = 0$ and $\Delta \phi_\text{end}$. 
This is ensured by comparing the 
CMB pivot scale $k_*^{-1}$ with today's Hubble horizon. With the expansion history of 
the universe from reheating until today described by the standard model of cosmology, 
this determines the number of e-folds $N_*$ of inflation elapsed after the scale $k_*^{-1}$ 
exited the horizon: $N_* = \ln ( a_0/a_*) + \ln(a_\text{end}/a_0)$. 
Tracking the expansion history of the universe epoch by epoch, 
this yields~\cite{Liddle:2003as,Martin:2010kz}: 
\begin{equation}
N_*^\text{cond} \simeq   \; 66.64 - \ln \left(\frac{k_*}{a_0 H_0}\right)  
+  \frac{1}{4}\ln{\left( \frac{V_*}{\Mp^4}\right) } 
+  \frac{1}{4}\ln{\left( \frac{V_*}{V_{\text{end}}}\right) }
+ \frac{1-3\omega}{12(1+\omega)} 
\ln{\left(\frac{\rho_{\RH}}{V_{\text{end}}} \right)} \,,
\label{eq:nefolds}
\end{equation}
where $\omega$ denotes the equation of state during reheating, $\rho_\text{rh}$  is the energy 
density related to the reheating temperature and  $k_* = a_* H_* =  0.002 \, \text{Mpc}^{-1}$ is the 
Planck pivot scale. $a_0$ and $H_0$ are the scale factor and Hubble constant today, where we 
work in units with $a_0 = 1$.  The constant first term on the right-hand side is determined by the 
scale factor and Hubble parameter at matter-radiation equality ($a_\text{eq}, H_\text{eq}$) and 
the degrees of freedom in the thermal bath at $a_\text{eq}$ and at reheating, $g_*^\text{eq}$ 
and $g_*^\RH $:
\begin{equation}
 \ln \frac{a_\text{eq} H_\text{eq}}{a_0 H_0} - \frac{1}{4} \ln \frac{3 H^2_\text{eq}}{\Mp^2} 
 - \frac{1}{12} \ln \frac{g_*^\RH }{g_*^\text{eq}} = 66.64 \,.
\end{equation}
$H_\text{eq}$ and $a_\text{eq}$ are determined by the matter density today, 
$\Omega_m^0 = 0.3156$, the red-shift of matter-radiation equality $z_\text{eq} = 3395$ 
and the Hubble parameter today, $H_0 = h \, 100 \, \text{km} \, \text{s}^{-1} \text{Mpc}^{-1}$, 
with $h = 0.6727$~\cite{Planck:2015xua}. As above $g_*$ is set to the corresponding 
standard model values, $g_*^\RH  = 427/4$ and $g_*^\text{eq} = 43/11$.

In the Monte Carlo scan we will impose the condition that $N_*$ should lie within the interval 
\begin{equation}
N_*^\text{cond}- 3\le N_*\le N_*^\text{cond}+3\,,
\label{eq:ncond}
\end{equation}
where we use $\omega=0$ as argued in Sec.~\ref{sec:linktr}. The reheating
temperature is computed from Eq.~\eqref{eq:TRdim5dim6} and the corresponding inflaton mass is obtained
from $V_{\text{end}}$ according to $m_\phi^2 = \epsilon_\text{end} V_\text{end}/\Mp^2$. The latter 
corresponds to Eq.~\eqref{eq:inflatonmass} if a single field $\phi$ is responsible for driving inflation 
and reheating. If this is not the case, larger values of $m_\phi$ and hence $\rho_\RH$ are possible. 
However $N_*$ only depends very weakly on this quantity, $\Delta N_* = (\ln \Delta \rho_\RH)/12$. 
For simplicity we hence use the equality instead of the more general inequality in Eq.~\eqref{eq:inflatonmass} 
for evaluating $N_*$, and take the uncertainty to be covered by allowing $N_*$ to be in a finite range 
around $N_*^\text{cond}$, cf.\ Eq.~\eqref{eq:ncond}. On the contrary, the lower bounds on $V_\text{end}$ 
and $\TR$ (cf.\ Eq.~\eqref{eq:TRdim5dim6}) are based on the more general inequality in 
Eq.~\eqref{eq:inflatonmass}, hence the results we present in Sec.~\ref{sec:results} are absolute lower bounds.
If the operator of the inflaton decay is not specified we impose the conservative condition 
\begin{equation}
N_*^\text{cond}(\text{dim}\,6)- 3\le N_*\le N_*^\text{cond}(\text{dim}\,4)+3\,,
\label{eq:ncondgen}
\end{equation}
with $N_*^\text{cond}(\text{dim 6})$ referring to Eq.~\eqref{eq:nefolds} with $\rho_\RH$ expressed by 
means of the dimension six case in Eq.~\eqref{eq:TRdim5dim6}. Accordingly, 
$N_*^\text{cond}(\text{dim 4})$ corresponds to the dimension four (i.e.\ unsuppressed) case, 
i.e.\ removing $m_\phi/\Mp$ in Eq.~\eqref{eq:dim5op}. In both cases, we set the 
coupling constant $\lambda$ to one.

\subsection{Monte Carlo scan \label{scan}}

In this subsection we briefly summarize the computational steps of the Monte Carlo scan over the 
parameter space spanned by the model \eqref{V3}, i.e., the parameters $V_*, c_1, c_2, c_3, c_4$. 
These 
parameters can be expressed as analytic functions of the CMB observables
$A_s, n_s,r,\alpha_s,\kappa_s$ to leading order in the slow-roll
parameters (using \eqref{sr1}) in order to obtain an estimate of the parameter 
region of interest, which we use in the course of generating scan points as described below.
However, after the generation of a parameter point we only keep the parameters
$V_*, c_1, c_2, c_3, c_4$ and recompute the CMB observables 
using the full expressions shown in the Appendix.

For each generated point of the Monte Carlo scan we perform the following steps. We first randomly choose 
$A_s, n_s$ using a Gaussian probability distribution around their mean values (see Eq.~\eqref{eq_planck_obs}) 
with various widths up to 3 standard deviations as well as $r$ 
with a logarithmically flat distribution between 
$10^{-3}$ and 0.3. This determines $V_*$, $c_1$ and $c_2$ (at leading order in the slow-roll parameters).
For variation of the coefficients $c_3$ and $c_4$, determined by $\alpha_s$ and $\kappa_s$ 
at leading order in the slow-roll parameters, we 
follow three different approaches used for generating different sets of parameter points. 
\begin{enumerate}
\item
Randomly choosing $\alpha_s$ and $\kappa_s$ 
using a Gaussian probability distribution around their mean 
values (see Eq.~\eqref{eq_planck_obs}) with up to 2 standard deviations.
\item
Randomly choosing $c_3$, $c_4$ using a logarithmically flat probability distribution with different ranges 
between $10^{-1}$ and $10^{-15}$.
\item
Randomly choosing $c_3$, $c_4$ close to values that lead to a vanishing discriminant (and discriminant of 
the discriminant) of the potential (see Sec.~\ref{sec:disk_method} for a detailed discussion).
\end{enumerate}
The complete set of parameter points has been found to provide a sufficiently dense coverage of the 
region of large potential drops $V_*/V_{\text{end}}$ and therefore ensures the formation of
a sharp transition between 
forbidden and allowed points.
Note that the absolute density of scan points does not claim any physical meaning.
\begin{figure}[h!]
\centering
\setlength{\unitlength}{1\textwidth}
\begin{picture}(1,0.44)
{\includegraphics[scale=0.54]{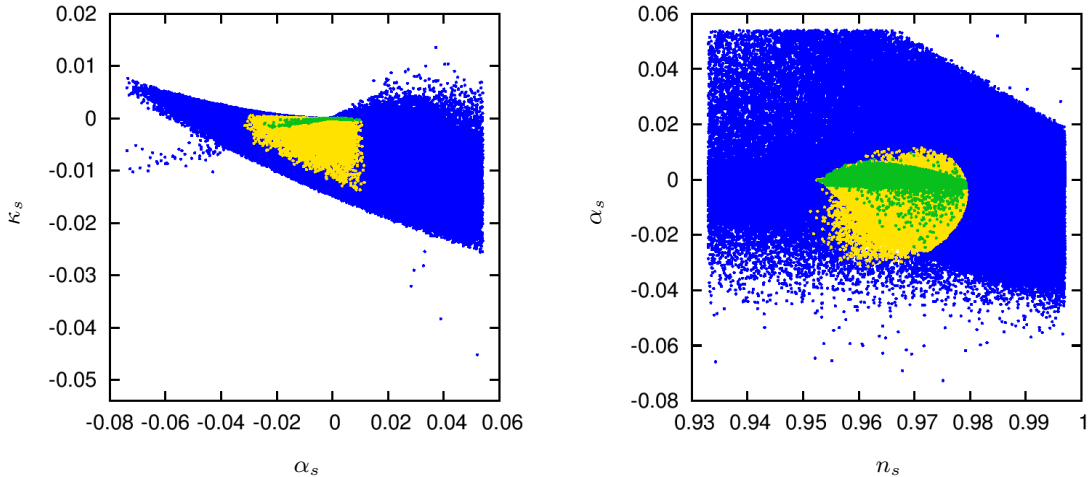}}
\end{picture}
\caption{
Scan points in terms of CMB observables. 
\emph{Left panel:}~Running ($\alpha_s$) and running of the running ($\kappa_s$) of the spectral index. 
\emph{Right panel:}~Spectral index $n_s$ and its running $\alpha_s$. \emph{Color-code:}~Blue:~Inflationary 
points. Yellow: Inflationary points within 2-sigma-contours regarding Planck data at $\Delta \phi = 0$. 
Green:~Points additionally fulfilling the constraint on $N_*$.
}
\label{fig:contri1}
\end{figure}

For each point we then compute $\Delta \phi_{\text{end}}$, requiring 
\begin{equation}
\epsilon(\Delta \phi_{\text{end}}) = \epsilon_{\text{end}}\,,
\label{eq:dphiepcon}
\end{equation}
where we use $\epsilon_{\text{end}} = 1$ (later we will also consider $\epsilon_\text{end} = 3$ for an estimation of the systematic uncertainties due to the use of the slow-roll approximation, cf.\ also footnote~\ref{ft_slowroll}). 
The solution of Eq.~\eqref{eq:dphiepcon} only corresponds to a physical solution, if $\Delta \phi_{\text{end}}<0$,
$V'(\Delta \phi_{\text{end}})>0$ and if there is no zero of $V'$ between 0 and $\epsilon_{\text{end}}$. 
Only points that fulfill {these} requirements  are kept as potential 
inflationary points. For each potential inflationary point we compute $N_*$ from
\begin{equation}
N_* = \int_{\Delta\phi_{\text{end}}}^{0} \frac{V}{V'}\frac{\D(\Delta\phi)}{\Mp^2}\,.
\end{equation}
Points with $10<N_*<10^5$ are kept. Further we compute the $\chi^2$ as described in 
Sec.~\ref{sec_planckdata} and impose the consistency condition for $N_*$ arising 
from the comparison of the CMB pivot scale with the present Hubble horizon as detailed in 
Sec.~\ref{sec:Nstarcon}. 
We generate a total of 15 million inflationary points. 

Fig.~\ref{fig:contri1} shows our scan points in the $\alpha_s$-$\kappa_s$ and $n_s$-$\alpha_s$ plane.
Inflationary points (i.e. potentials which support slow-roll inflation for $10<N_*<10^5$ e-folds) are 
marked in blue whilst points that additionally fulfill the Planck constraints on the scalar potential at 
$\Delta \phi = 0$, as described in Sec.~\ref{sec_planckdata},  are marked in yellow. Note that these 
points do not represent potentials fully compatible with the Planck data, as the Planck data 
additionally contains the information that $\Delta \phi = 0$ corresponds to a pivot scale of 
$k_* = 0.002 \text{ Mpc}^{-1}$. Rather these yellow points (referred to as ``locally Planck allowed'' 
in the following) represent potentials which are locally consistent with the Planck data around 
$\Delta \phi = 0$, but in general will not give the observed amount of inflation between 
$\Delta \phi = 0$ and $\Delta \phi_\text{end}$. This final condition is imposed by means of the 
consistency condition Eq.~\eqref{eq:ncondgen} and results in the points marked in green.
As mentioned in Sec.~\ref{sec_planckdata} points that fulfill Eq.~\eqref{eq:ncondgen} only 
survive in a very small interval in $\kappa_s$. 
The stray blue points around $\alpha_s \sim 0$, $\kappa_s < -0.02$ in the left panel correspond 
to very small values of $V_*/V_\text{end}$ and low values of $N_*$, close to the lower bound 
$N_* > 10$. Since this region of the parameter space is irrelevant for the lower bounds on 
$V_\text{end}$ and $T_\RH$, the density of scan points is very low here.

\subsection{Properties of the inflaton potential \label{sec:disk_method}}

\begin{figure}
\centering
\setlength{\unitlength}{1\textwidth}
\begin{picture}(0.75,0.42)
{\includegraphics[scale=0.54]{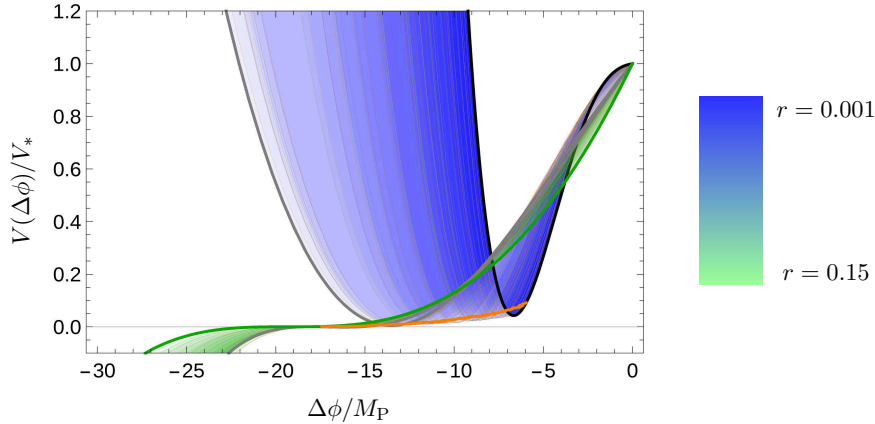}}
\end{picture}
\caption{Scalar potentials which maximize the energy drop during the final $N_*$ e-folds of 
inflation for different values of the tensor-to-scalar ratio $r$ (color-coded on a logarithmic scale). 
The orange line marks the end of inflation at $\epsilon = 1$. }
\label{fig_potentials}
\end{figure}

The parameter region of particular interest to this study is the regime which allows for the largest 
decrease of the potential energy during inflation. This implies a fairly steep potential allowing for 
a large energy drop during the final $N_*$ e-folds while at the same time keeping the slow-roll 
parameters sufficiently small. In particular, not only must the derivatives of the potential be small 
compared to the inflationary scale $V_*$, but they must `track' the rapidly falling energy density 
$V(\Delta \phi)$ until after $N_*$ e-folds the slow-roll parameter $\epsilon$ reaches its critical 
value and inflation comes to an end. In Fig.~\ref{fig_potentials} we schematically show the type 
of potentials which meet these requirements, as found in the Monte Carlo scan. Here, we have 
imposed the 95$\%$ confidence interval from the CMB data, cf.\ Sec.~\ref{sec_planckdata}, and 
required values of $N_*$ in accordance with Eq.~\eqref{eq:ncondgen}. The color-coding refers 
to the tensor-to-scalar ratio $r$. We immediately notice that we can identify two different regimes, 
corresponding to potentials with minimum (blue) or saddle point (green) close to the end of inflation.

This can be understood as follows. As $V(\Delta \phi)$ decreases towards the end of inflation, 
the slow-roll parameter $\epsilon$ increases and inflation comes to an end -- unless the 
numerator $V'(\Delta \phi)$ of $\epsilon$ vanishes simultaneously. The smallest values for 
$V_\text{end}$ are hence achieved if $V(\Delta \phi)$ and $V'(\Delta \phi)$ have a common 
zero point at a suitable value of $\Delta \phi$. This implies an (at least) double zero point of 
$V(\Delta \phi)$ and hence a necessary condition for the smallest $V_\text{end}$ is a 
vanishing discriminant of $V(\Delta \phi)$. In practice, small values of $V_\text{end}$ after 
a finite amount of e-folds are achieved if this condition is approximately satisfied, i.e.\ if the 
zero-points approximately coincide,
\begin{equation}
 \text{Disc}[V(\Delta \phi)] \simeq 0 \,.
\label{eq_disk}
\end{equation}
 Eq.~\eqref{eq_disk} is an equation of degree three for the parameter $c_4$ with generically 
 one real and two complex solutions. We observe that potentials with particularly small 
 $V_\text{end}$ are achieved if (at least) two of these solutions approximately coincide, i.e.\
 \begin{equation}
 \text{Disc}[\text{Disc}[V(\Delta \phi)](c_4)] \simeq 0 \,.
\label{eq_diskdisk}
\end{equation}
This condition implies that $\text{Disc}[V(\Delta \phi)]$ has a double zero point at 
$c_4^0 \in \mathbb{R}$, ensuring that the smallness of $\text{Disc}[V(\Delta \phi)]$ 
is stable against small variations of the parameter $c_4$, allowing us to find solutions 
for $V(\Delta \phi)$ with (several) close, but not exactly coinciding zero-points -- precisely 
what we are after.
The two distinct situations observed in Fig.~\ref{fig_potentials} correspond to potentials 
with two (blue) or three (green) almost coinciding zero-points. Four coinciding zero-points 
would simply correspond to a $\phi^4$ potential, which is however clearly excluded by the Planck data.
Eqs.~\eqref{eq_disk} and \eqref{eq_diskdisk} provide us with analytical solutions for two 
parameters ($c_4$ and $c_3$). Using values in the close vicinity of these solutions significantly 
increases the efficiency in exploring the large $V_*/V_\text{end}$ region. 

Moreover, we can exploit that certain ranges of the coefficients $c_n$ lead to particularly large 
energy drops during inflation. For example, small values of the higher coefficients lead to 
well-behaved potentials, which are more likely to support inflation over a sufficient amount 
of e-folds. Also, we note that for small values of $r$, the $\Delta \phi^3$-coefficient is indirectly 
proportional to the square root of the tensor-to-scalar ratio, $c_3 \propto -\alpha_s/\sqrt{r} \propto 1/c_1$. 
This correlation is only strengthened by imposing the requirement on $N_*$. Implementing these 
observations allows us to efficiently investigate the region of large $V_*/V_\text{end}$.

\section{Results and discussion}\label{sec:results}

In this section, we present and discuss the results of the Monte Carlo scan. Following the 
strategy of the analytical discussion, cf.\ Sec.~\ref{sec:2}, we will first discuss the constraints 
on the scalar potential during the final $N_*$ e-folds before drawing conclusions about the 
following reheating process.

\subsection{The energy density at the end of inflation}

\begin{figure}[h!]
\centering
\setlength{\unitlength}{1\textwidth}
\begin{picture}(1,0.44)
{\includegraphics[scale=0.66]{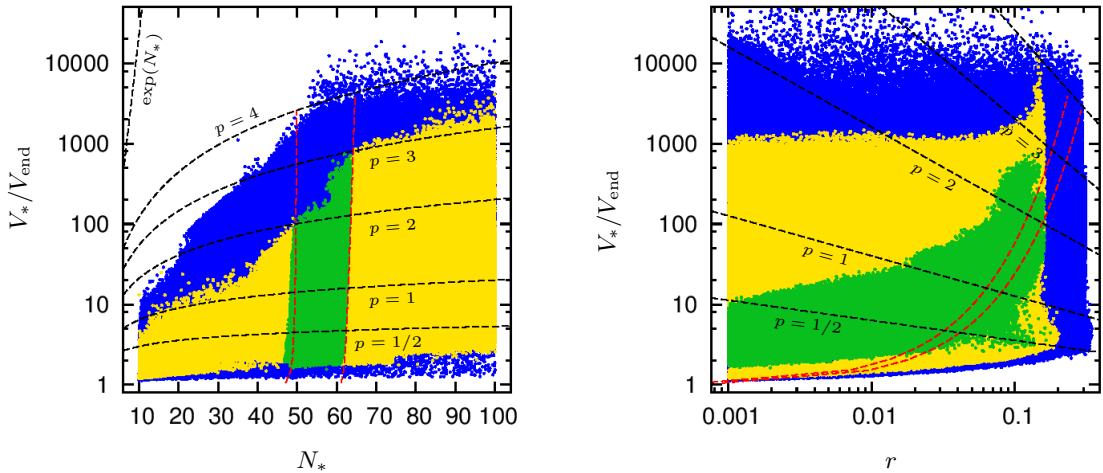}}
\end{picture}
\caption{
Scan points in the $N_*$-$(V_*/V_{\text{end}})$-plane and $r$-$(V_*/V_{\text{end}})$-plane.
\emph{Color-code:}~Blue:~Inflationary points. Yellow: Inflationary points within 2-sigma-contours
regarding Planck data at $\Delta \phi = 0$. Green:~Points additionally fulfilling the constraint on $N_*$. 
Dashed black curves:~Prediction for monomial potentials $\phi^p$ with $p = \{1/2, 1, 2, 3, 4 \}$ with the 
constraint on $N_*$ fulfilled within the region marked dashed red (for $p \leq 4$). In the left panel, we 
additionally show the most conservative analytical bound, the exponential potential.
}
\label{fig:contri2}
\end{figure}

The dynamics of inflation, described by Eq.~\eqref{eq_Nstar}, is independent of the overall scale 
$V_*$ of the potential as defined in Eq.~\eqref{V3}. As a first step, let us hence consider the relative 
drop of the scalar potential during inflation, $V_*/V_{\text{end}}$, as a function of $N_*$ and $r$, 
respectively, cf.\ Fig.~\ref{fig:contri2}. Here we mark in blue all parameter points which sustain 
inflation for $N_* = 10 \,\text{-}\, 10^5$  e-folds -- implying that these potentials also contain a 
mechanism to end inflation by reaching $\epsilon(\Delta\phi) = \epsilon_\text{end} = 1$.\footnote{In 
this sense, we restrict our analysis to single-field inflation models, excluding in particular models 
of hybrid inflation where the dynamics of a second scalar field is responsible for ending inflation.} 
In yellow, we show the parameter points which additionally locally match the Planck data at 
$95\%$ C.L., cf.\ Sec.~\ref{sec_planckdata}. Finally in green, we impose the consistency 
condition on $N_*$ given  by Eq.~\eqref{eq:ncondgen}. For comparison, the black dashed 
lines show the results obtained in Sec.~\ref{sec:2} for the monomial potentials $V \propto \phi^p$ 
and the exponential potential $V \propto \exp(\sqrt{2 \epsilon_\text{end}} \Delta \phi/\Mp)$, 
cf.\ Eqs.~\eqref{eq:expbound} and \eqref{eq:Vratmon}. For monomials with $p \leq 4$, the red 
dashed lines indicate the range of $N_*$-values in accordance with Eq.~\eqref{eq:ncondgen}.

Focusing on the locally Planck allowed points in the left panel, we first note that all points 
covered by our ansatz and locally allowed by the Planck data are far below the absolute upper bound 
for $V_*/V_\text{end}$ given by the exponential potential. Moreover we can classify these points 
by the number of approximately coinciding zero-points at $\Delta\phi \sim \Delta\phi_\text{end}$, 
cf.\ Sec.~\ref{sec:disk_method}.
Points with one, two or three approximately coinciding zero-points are predominantly located in 
three horizontal bands yielding increasing values for $V_*/V_\text{end}$, roughly bounded from 
below by $V_*/V_\text{end} = \{1, \, 25,  \, 250 \}$, respectively, and broadening for increasing 
values of $N_*$. In the lowest band $V(\Delta \phi)$ has a single (approximate) zero-point. As 
$\Delta \phi$ approaches this value, the slow-roll parameter $\epsilon$ blows up rapidly and 
inflation ends. The higher bands correspond to two and three (approximately) coinciding 
zero-points, i.e.\ common zero points in $V(\Delta \phi)$ and $V'(\Delta \phi)$, allowing 
$\epsilon$ to keep a finite value as $V(\Delta \phi)$ approaches very small values.
The monomial potentials with  $p = \{1,2,3\}$ can be seen as prototypical examples.

Also in the $r$-$V_*/V_\text{end}$ plane, the points locally allowed by the Planck constraint can be 
grouped according to number of coinciding zero-points, yielding three horizontal bands 
which turn upwards at $r \sim 0.1 \,\text{-}\, 0.2$. The structure of the resulting yellow region 
in the right panel of Fig.~\ref{fig:contri2} can be understood based on Eq.~\eqref{eq_disk}. 
Taking $\alpha_s$ and $\kappa_s$ to be close to zero as indicated by the Planck data and 
$n_s$ within the $2 \sigma$ region, Disc$[V(\Delta \phi), \Delta \phi]$ is approximately 
constant for small $r$ until reaching a zero-point around $r \sim 0.11 \,\text{-}\, 0.23$. In 
this region of large $r$, very small values of Disc$[V(\Delta \phi), \Delta\phi]$ are reached, 
corresponding to (several) nearly coinciding zero-points allowing for large $V_*/V_\text{end}$. 
In this sense, the recent Planck results have now reached a crucial sensitivity, excluding the 
part of the parameter space where $V_*/V_\text{end}$ begins to shoot up. Correspondingly, 
if future measurements are able to constrain the tensor-to-scalar ratio to be below about 
$r \sim 0.05$, the upper bound on $V_*/V_\text{end}$ will become about an order of 
magnitude stricter. The overall shape of the green region can be understood by 
employing monomial potentials as prototypical examples. High values of $V_*/V_\text{end}$ 
are reached if $V(\Delta \phi)$ has a multiple zero-point ($p \geq 2$, corresponding to a 
sizable value of $r$) whereas at small values of $r$, for $p \ll 1$, the potential becomes 
very flat and $V_*/V_\text{end} \rightarrow 1$.

\begin{figure}[h!]
\centering
\setlength{\unitlength}{1\textwidth}
\begin{picture}(1,0.44)
{\includegraphics[scale=0.66]{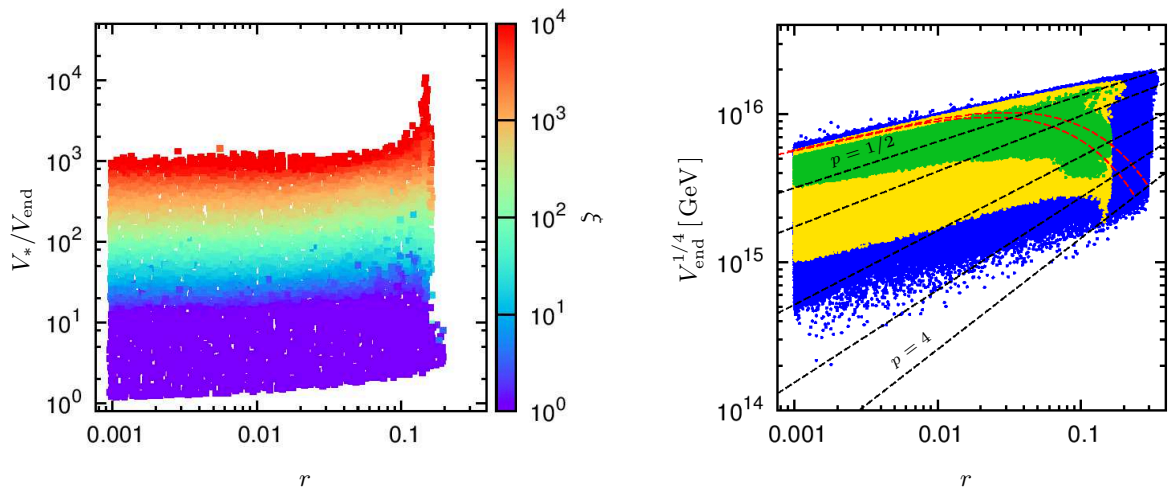}}
\end{picture}
\caption{
Fine-tuning $\xi$ among the locally Planck allowed points in the $r$-$(V_*/V_{\text{end}})$-plane (left) and 
lower bound on the energy density at the end of inflation in the $r$-$V_{\text{end}}$-plane (right). 
The color-code in the right panel is as in Fig.~\ref{fig:contri2}.
}
\label{fig:contri3}
\end{figure}

From the discussion above we can anticipate that the parameter points with high 
$V_*/V_\text{end}$ come with finely tuned parameters. To quantify this, we determine 
the fine-tuning measure~\cite{Barbieri:1987fn} 
\begin{equation}
 \xi \equiv 
 \text{Max} \left[ \frac{\delta (V_*/V_\text{end})}{\delta c_n} \frac{c_n }{(V_*/V_\text{end})}\,,\; {n = 1,\dots,4} \right] \,,
\end{equation}
with $\delta c_n$ denoting a small variation of the respective parameter of the scalar potential~\eqref{V3}, 
$|\delta c_n/c_n| \sim 10^{-6}$. Large values of $\xi$ indicate a high degree of fine-tuning, implying 
that the value of $V_*/V_\text{end}$ is extremely sensitive to at least one of the parameters. On the other hand, 
points with $\xi = {\cal O}(1 \,\text{-}\,  10)$ correspond to potentials which do not require any particular tuning. 
In the left panel of Fig.~\ref{fig:contri3} we show the fine-tuning measure $\xi$ 
where we performed a binning among the locally Planck allowed points in the $r$-$(V_*/V_\text{end})$ plane and
displayed the point with the smallest value of $\xi$ in each bin.
As expected, high values of $V_*/V_\text{end}$ always come with a high degree of tuning. 
This is rather independent of the value of $r$. In particular, the three horizontal `bands' mentioned above,
corresponding to an increasing number of coinciding zero-points, are also clearly distinguished by 
increasingly higher fine-tuning. Dropping the requirement of lying within the $2$-sigma-region of the 
Planck data and  considering the full data set, we observe that the fine-tuning is relaxed for small $r$ 
by up to a factor of 10 but remains basically unchanged for $r \sim 0.1$. 
 The high tuning is hence mainly a result of the careful balance of $V$ and $V'$ in the slow-roll parameter 
 $\epsilon$, and is not a feature of the specific values of the CMB observables, in particular for large 
values of $r$. 
Comparing the left panel of Fig.~\ref{fig:contri3} with the points passing all constraints in the  right 
panel of Fig.~\ref{fig:contri2}, we note that in particular the large values of $V_*/V_\text{end}$ around 
$r \sim 0.1$ feature a maybe uncomfortably high degree of tuning. A restriction to less severely tuned 
potential would correspondingly yield a tighter upper bound on $V_*/V_\text{end}$, translating to a 
tighter lower bound on the reheating temperature. Since choosing the acceptable degree of tuning 
is however a very subjective question, we continue in the following with the full data set.

Finally, in the right panel of Fig.~\ref{fig:contri3}, we include the information on $V_*$ which comes 
with an additional $r$-dependence. This leads to the final constraints on the energy density at the 
end of inflation, one of the main results of this paper. Independent of any assumptions about the 
reheating process which we will make in the following section, we find that the CMB observations 
constrain this energy scale to $3 \times 10^{15} \, \GEV < (V_\text{end})^{1/4} < 2 \times 10^{16} \, \GEV $ 
for $r > 10^{-3}$ for the scalar potential of any single-field inflation model described by operators 
up to dimension four. This remarkable narrow range close the GUT scale might be taken as an 
indication to explore connections between inflation and the breaking of a unified gauge group. 
While this has been recently much discussed in the context of a large tensor-to-scalar ratio, we 
stress that indeed this conclusion holds over the entire range of the tensor-to-scalar ratios investigated, 
$0.001 < r < 0.15$, since the overall scaling of the potential, $V_* \propto r$, is to some extent 
compensated by the large decrease in energy allowed by large $r$. This is illustrated by the 
example of the monomial potentials, cf.\ dashed red lines in Fig.~\ref{fig:contri3}. For small values of 
$r$, the drop in energy during inflation becomes negligible and the energy at the end of inflation is 
mainly determined by the $r$-dependence of $V_*$. As an aside, we point out that this provides a 
good indication on how to extend our bounds to smaller values of $r < 10^{-3}$.  The results depicted 
in the right panel of Fig.~\ref{fig:contri3} stress the importance of upcoming CMB experiments. 
The detection of a tensor-to-scalar ratio in the percent regime, $0.01 \leq r \leq 0.06$, would tighten 
the bound on the energy density at the end of inflation to $(V_\text{end})^{1/4} > 5 \times 10^{15}$~GeV.

Bounds on the scalar potential of inflation have been derived in other contexts. Using the same 
ansatz as Eq.~\eqref{V3}, the Planck collaboration recently performed a local reconstruction of 
the scalar potential around $\phi_*$, restricting themselves to the observable part of the 
potential~\cite{Ade:2015lrj}. We find that our ranges for the parameters $c_n$ lie well 
within the ranges found in this local reconstruction. This is an important cross-check, 
implying that also a few e-folds before the pivot-scale our potentials are sufficiently well 
behaved to not generate any anomalies at low multipoles in the CMB. Compared to 
Ref.~\cite{Ade:2015lrj}, our parameter ranges are of course much more constrained 
since we additionally impose the requirement that extending Eq.~\eqref{V3} to large 
values of $\Delta \phi$, the scalar potential supports slow-roll inflation for a suitable 
amount of e-folds. Moreover, recently in Ref.~\cite{Antusch:2014saa} a  lower bound 
on $V_*/V_\text{end}$ was derived, $V_*/V_\text{end} > 1.01$ for $r < 0.15$, in this 
sense focussing on the opposite limit as the bound presented here. 

\subsection{A lower bound on the reheating temperature \label{sec:Trhbound}}

\begin{figure}[h!]
\centering
\setlength{\unitlength}{1\textwidth}
\begin{picture}(1,0.44)
{\includegraphics[scale=0.66]{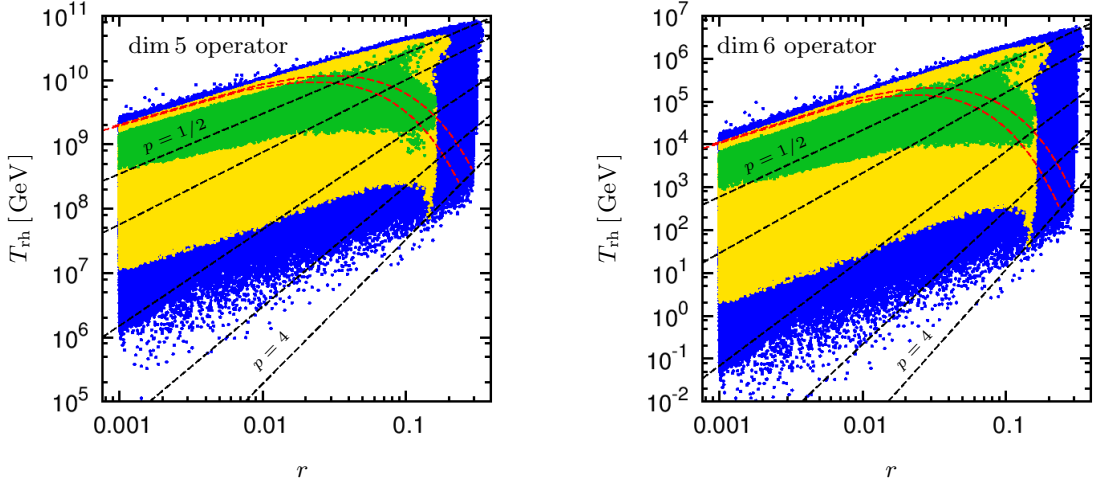}}
\end{picture}
\caption{
Scan points in the $r$-$\TR$-plane showing the lower bound on the reheating temperature
for a decay via dimension five operators (left) 
and dimension six operator (right) with $\lambda = 1$. The color-code is as in Fig.~\ref{fig:contri2}.
}
\label{fig:contri4}
\end{figure}

Equipped with a lower bound on the energy scale at the end of inflation, $V_\text{end}$, 
we can proceed to constrain the reheating temperature based on Eqs.~\eqref{eq:inflatonmass} 
and \eqref{eq:TRdim5dim6}. In Fig.~\ref{fig:contri4} we show the resulting lower bound on $\TR$ 
based on a dimension five and dimension six operator for the inflaton decay, respectively. 
Based on the monomials as prototypical examples the overall scale and shape of this 
constraint is well understood by applying Eq.~\eqref{Trh_mon}. 
In both cases, we obtain values significantly above the BBN constraint of about 10~MeV,
\begin{equation}
 \begin{split}
   T_\RH^\text{dim 5} & > 3 \times 10^8~\GEV \,,  \\ 
T_\RH^\text{dim 6} & > 7 \times 10^2~\GEV  \,. 
\label{eq_Trhbounds}
 \end{split}
\end{equation}
for $r > 0.001$.
Here, as in Fig.~\ref{fig:contri4}, we have set the coupling constant in Eq.~\eqref{eq:TRdim5dim6} to 
$\lambda = 1$. Note that $T_\RH$ depends linearly on $\lambda$, hence a suppression of this 
coupling constant can weaken the bound accordingly.

The left panel of Fig.~\ref{fig:contri4} displays a remarkably strong bound for the dimension 
five case, implying that temperatures at least very close to the temperatures  relevant for the 
gravitino problem and thermal leptogenesis were indeed reached in the early universe 
($T \sim 10^9$~GeV). This bound might be further tightened if upcoming B-mode searches 
succeed in detecting a tensor-to-scalar ratio of a few percent, $0.01 < r < 0.07$. In this region, 
the lower bound on the reheating temperature for $\lambda = 1$ reaches as high as 
$T_\RH > (1\,\text{-}\,2) \times 10^9$~GeV. This bound is however significantly weakened 
if the coupling $\lambda$ is suppressed, either by a symmetry or by an accidentally small 
value. To demonstrate this effect, we consider the situation where the dimension five 
operator is negligibly small and the time-scale of reheating is instead governed by a 
dimension six operator in the right panel of Fig.~\ref{fig:contri4}. In this case the 
constraint on the reheating temperature is significantly weaker, owing to the additional 
factor of $m_\phi/\Mp \sim \sqrt{V_\text{end}}/\Mp^2$, but is still well above the BBN 
bound of 10~MeV.

Throughout this study we have employed $\epsilon = 1$ as the condition for the end of inflation. 
Since slow-roll is violated at the end of inflation, this is however only approximately correct, 
more precisely inflation ends at $\epsilon_H = \dot \phi^2/(2 H^2) = 1$~\cite{Liddle:1994dx}. 
The effect of this on the bounds derived on the quantities of the reheating process is 
discussed in footnote~\ref{ft_slowroll}, here we comment on the resulting uncertainty 
of the quantities $V_*/V_\text{end}$ and $T_\RH$ due to a misestimation of $\Delta \phi_\text{end}$. 
To this end we have solved the full second order differential equation~\eqref{eq_diff2} for 
a representative subset of 20000 parameter points.\footnote{To obtain robust statements, 
we consider three different parameter sets: After binning our points in the $r$-$(V_*/V_\text{end})$ 
plane, we consider first a randomly chosen point within each bin and, second, the point with the highest 
and lowest fine-tuning, respectively. Third, we re-compute our results for $\epsilon_\text{end} = 3$ 
(which is found to be a typical value for $\epsilon(\epsilon_H = 1)$), and select the allowed points 
with highest $V_*/V_\text{end}$. \smallskip} (Unfortunately, applying this procedure in the full 
parameter scan is computationally too demanding.) As expected, we find 
$\epsilon(\epsilon_H = 1) \gtrsim 1$, hence the slow-roll approximation underestimates 
$\Delta \phi_\text{end}$ and thus $V_*/V_\text{end}$. We find that without the slow-roll 
approximation, $N_*$ increases by $\Delta N_* = 1.1\;$-$\;2.2$ and $V_*/V_\text{end}$ 
increases by an ${\cal O}(1)$ factor, which in the most interesting range $45 < N_* < 65$ 
is found to be  $(V_*/V_\text{end})_\text{full}/(V_*/V_\text{end})_\text{sr} \simeq 1.8\;$-$\;2.1$. 
We observe that since $\epsilon$ grows very rapidly at the end of inflation in the models of 
interest here, the errors in $\Delta \phi_\text{end}$ and the derived quantities are small.
The bound~\eqref{eq:inflatonmass} on $m_\phi$, entering the bounds on the reheating 
temperature via Eq.~\eqref{eq:TRdim5dim6}, changes by a factor of $0.7\;$-$\;1.8$. 
This moderate change is due to compensating effects of $\epsilon_\text{end}$, which 
increases, and $V_\text{end}$, which decreases  with respect to the slow-roll approximation.
In summary the uncertainty arising from the use of the slow-roll approximation is very moderate.

In the analysis presented here, we have restricted ourselves to a matter-dominated reheating 
phase, $\omega = 0$, while allowing for a wide range of inflationary potentials.\footnote{Note 
that as argued around Eq.~\eqref{eq:mbound2}, we expect the dependence of our analysis 
on the value $\omega$ to be small.} An alternative approach to constrain the reheating 
temperature is to fix the inflation model and then use Eq.~\eqref{eq:nefolds} to constrain 
the reheating temperature. This allows one to probe all reheating scenarios with a constant 
$\omega$. A recent analysis along these lines can be found in Ref.~\cite{Cook:2015vqa}. 
For simple inflationary models, such as the monomial potentials which have served as 
prototypical examples throughout this paper, the resulting bounds on the reheating 
temperature can be determined analytically. For a $\phi^2$ monomial potential and 
$\omega = 0$, the reheating temperature is found to be larger than about $10^9$~GeV 
from the $1\sigma$ constraints on the spectral index $n_s$, consistent with our results. 
However, lowering the power of the monomial to $p = 1$, the bound completely disappears, 
allowing for reheating temperatures in the full range of $10~\MEV < T_\RH < 2 \times 10^{16}~\GEV$. 
The result is also very sensitive to the choice of $\omega$,  increasing its value to $\omega = 2/3$
implies for the $p=2$ case that no significant bounds can be placed on the reheating temperature, 
whereas for $\omega = -1/3$ the lower bound is raised to about $10^{14}$~GeV. While it is 
remarkable that the CMB data alone can yield strong constraints on the reheating temperature 
for certain inflation models, these results also stress the need for a more model independent 
approach, as we have pursued in this paper.

Further approaches to constrain the reheating temperature include constraints from the 
thermalization of neutrinos, measured by the effective number neutrinos $N_\text{eff}$ 
required to reproduce the CMB spectrum~\cite{DeBernardis:1900zz, Ichikawa:2005vw}, 
implying however only a rather weak bound of $T_\RH > 3.2~\MEV$. Moreover, a detection 
of gravitational waves by future space-based experiments such as BBO~\cite{Crowder:2005nr} 
and DECIGO~\cite{Kawamura:2011zz} would open a completely new window to the energy 
scales of the early universe. Given a sufficiently large tensor-to-scalar ratio, this would allow 
to probe the reheating in a range of $10^6 \,\text{-}\, 10^9~\GEV$~\cite{Nakayama:2008wy}. 
Finally, from a theoretical point of view it has been argued that there is an upper bound on 
the reheating temperature, $T < 10^9~\GEV$, since large couplings of the inflaton field to 
any other particles would destroy the flatness of the inflaton potential via radiative 
corrections~\cite{Kofman:1996mv}. This bound can however easily be avoided if the 
time-scale of reheating is set by the decay of a particle different than the inflaton, as 
can occur e.g.\ in hybrid inflation or in some preheating scenarios.

\section{Conclusion \label{sec:conclusion}}

In this paper we derived constraints on the relevant energy scales of the very early universe,
namely the energy scale at the end of inflation, $V_{\text{end}}$, the mass of the
inflaton, $m_{\phi}$, and the reheating temperature, $\TR$. The most
dramatic potential drop $V_*/V_{\text{end}}$ is realized in an exponential potential yielding
the most conservative bound $V_*/V_\text{end} < e^{N_*}$. However, for a realistic single-field inflation
scenario -- providing a mechanism to end inflation as well as being consistent with observations
of the CMB around the pivot scale -- a much stronger bound can be achieved. In order to cover a
large class of relevant inflation models we examined a general fourth order polynomial potential.
We constrained this potential by requiring consistency with recent data from the Planck collaboration 
and by demanding that inflation ends after the number of e-folds which provides a matching between 
the CMB pivot scale and today's Hubble horizon. We found that this restricts the potential drop to stay
below $V_*/V_{\text{end}}\simeq1000$, yielding the highest values for large 
tensor-to-scalar-ratios close to being excluded by Planck, $r\simeq0.1$. For smaller $r$ the potential 
drop becomes significantly smaller. This results in the bound $V_{\text{end}}^{1/4}>3\times 10^{15}\GEV$
for $r>0.001$ -- remarkably close to the GUT scale.

Matching the inflaton potential at the end of inflation to the subsequent oscillation phase allowed us to 
connect $V_{\text{end}}$ and $m_{\phi}$ which provides a lower bound on the inflaton mass.
This allows us in turn to estimate the decay width of the inflaton governing the time scale of the
energy transition to the thermal bath and hence the energy density at the beginning of the radiation 
dominated phase. Without taking into account the decay width of the inflaton, the reheating temperature
can only be constrained by solving Eq.~\eqref{eq:nefolds} for $\rho_{\RH}$. However, due to the logarithmic
dependence this procedure is subject to huge uncertainties, in particular it does not provide an interesting
bound in the general class of models considered in this paper. Instead we specified the decay mode
of the inflaton by exemplarily considering generic Planck-suppressed dimension five and six operators. 
This allowed us to place bounds on the reheating temperature, 
$T_\RH^\text{dim\,5} > 3 \times 10^8~\GEV$ and $T_\RH^\text{dim\,6} > 7 \times 10^2~\GEV$, significantly
above the requirement that thermalization of the universe has to take place well before BBN, $\TR>10\MEV$. 
If moreover upcoming B-mode searches succeed in detecting a tensor-to-scalar ratio of a few percent, 
$0.01 < r < 0.07$, our results imply a particularly interesting bound in the dimension five case, 
$\TR > 10^9$~GeV. For a reheating temperature in this range,  the gravitino problem provides 
very challenging limitation on the supersymmetric parameter space. 

We emphasis that our results are robust in various concerns. 
The derived bounds do not depend strongly on the choice of $\epsilon_\text{end}$. Although our
results are achieved for single-field inflation models we expect the bounds to conservatively hold for 
the situation in which inflation is ended by a second field and hence $\epsilon_\text{end}$ is not required to
become large towards the end of inflation.  
In order to match the inflaton potential at the end of inflation to the subsequent oscillation phase
we assumed a (predominantly) quadratic potential in the latter phase, i.e.\ an equation of state 
$\omega=0$. We argued that even for a different choice of $\omega$ our results are not changed
significantly.
Throughout this study we worked in the slow-roll approximation. 
Nevertheless we have discussed that our bounds are expected to hold up to an 
$\mathcal{O}(1)$ factor even
if the slow-roll conditions are
slightly violated. 
In contrast to the robustness of the bound on $V_{\text{end}}$ there, however, remains a huge dependence of the
bound on $\TR$ to an unknown quantity, namely the type and coefficient of the operator governing the
inflaton decay. This amounts in a difference of almost six orders of magnitude between the case of a
dimension five and six operator.

\vspace{1.2cm}
\subsubsection*{Acknowledgements}

We thank Stefan Antusch, Marco Drewes, Julien Lesgourgues, David Nolde, Martin Schmaltz 
and Alexander Westphal for very helpful discussions. 
This work was supported by the European Union Program FP7 ITN INVISIBLES  
(Marie  Curie  Actions,  PITN-GA-2011-289442).


\section*{Appendix}
\begin{appendix}

\section{CMB observables beyond leading order}\label{app:CMBobsN}

In this Appendix we present the full expressions for the CMB observables in 
terms of the potential slow-roll parameters used
for the computation of $\chi^2$ as described in Sec.~\ref{sec_planckdata}.
In all expressions the slow-roll parameters are evaluate at $\phi=\phi_*$.
Note that there does not exist a slow-roll parameter beyond $\sigma^3$
for the considered class of potentials \eqref{V3} as the fifth derivative 
of the potential vanishes.
To second order in the slow-roll parameters the scalar amplitude reads~\cite{Gong:2001he} 
\begin{equation}
 \begin{split}
A_s  =\;
& \frac{V}{24 \pi ^2 \epsilon\,\Mp^4}\left[1+ \left(-\frac{1}{3}-6 C \right)\epsilon +\left(-\frac{2}{3}+2 C \right) \eta \;+ 
\right. 
\\
&
\left(-\frac{154}{3}+\frac{11 \pi ^2}{2}-\frac{4 C }{3}+6 C ^2 \right)\epsilon^2+
\left(\frac{284}{9}-\frac{11 \pi ^2}{3}+\frac{4 C }{3}-4 C ^2\right) \epsilon  \eta \;+
\\
&
\left.
\left(-\frac{37}{9}+\frac{\pi ^2}{2}-\frac{2 C }{3}+2 C ^2\right) \eta ^2+
\left(-\frac{2}{9}+\frac{\pi ^2}{12}+\frac{2 C }{3}-C ^2\right) \xi^2\right]\,.
 \end{split}
\end{equation}
In the above expressions we introduced
\begin{equation}
C = -2 + \ln 2 + \gamma_\text{E}\simeq-0.729637\,,
\end{equation}
where $\gamma_\text{E}$ is the Euler-Mascheroni constant, $\gamma_\text{E}\simeq 0.577216$.
From $A_s$ we can compute
\begin{align}
 n_s -1&= \frac{\D \ln A_s}{\D \ln k}\,, \\
 \alpha_s &=  \frac{\D n_s}{\D \ln k}\,, \\ 
 \kappa_s &= \frac{\D \alpha_s}{\D \ln k}\,,
\end{align}
using the general relation~\cite{Gong:2001he}
\begin{equation}
\frac{\D X|_{aH=k}}{\D \ln k}
=- \left. \left( 1 + \frac{1}{3}\epsilon + \frac{1}{3}\eta
+ \frac{5}{9} \epsilon^2 - \frac{4}{9} \epsilon\eta
+ \frac{2}{9}\eta^2 + \frac{1}{9}\xi^2 +\mathcal{O}\left(\epsilon^3\right) \right)
\frac{V'}{V}\frac{\D X}{\D \phi} \right|_{aH=k}\,.
\label{eq:devSLpot}
\end{equation}
We obtain
\begin{equation}
 \begin{split}
 n_s = \;&
1-6 \epsilon+2 \eta +\left(-\frac{10}{3}-24 C\right) \epsilon^2-(2-16 C) \epsilon  \eta 
+\frac{2 }{3}\eta^2+\left(\frac{2}{3}-2 C\right) \xi ^2
+ \\ &\left(-\frac{3734}{9}-\frac{104 C}{3}-96 C^2+44 \pi ^2\right) \epsilon ^3  
+\left(\frac{1190}{3}-\frac{4 C}{3}+96 C^2-44 \pi ^2\right) \epsilon ^2 \eta 
+ \\ &\left(-\frac{742}{9}+12 C-16 C^2+\frac{28 \pi ^2}{3}\right) \epsilon  \eta ^2
+\frac{4}{9}\eta ^3
+\left(-\frac{98}{3}+4 C-12 C^2+4 \pi ^2\right) \epsilon  \xi ^2
+ \\ &\left(\frac{28}{3}- \frac{8 C}{3}+C^2-\frac{13 \pi ^2}{12}\right) \eta  \xi ^2
+ \left(\frac{2}{9}-\frac{2 C}{3}+C^2-\frac{\pi ^2}{12}\right) \sigma ^3\,,
 \end{split}
\end{equation}
\begin{equation}
 \begin{split}
 \alpha_s=\;&
-24 \epsilon ^2+16 \epsilon  \eta -2 \xi ^2 
- \\ &
\left(\frac{104}{3}+192 C\right) \epsilon ^3-\left(\frac{4}{3}-192 C\right) \epsilon ^2 \eta 
+4 (3-8 C) \epsilon  \eta ^2 
+(4-24 C) \epsilon  \xi ^2- \\ & \left(\frac{8}{3}-2 C\right) \eta  \xi ^2-\left(\frac{2}{3}-2 C\right) \sigma ^3
-  \\ & \left(+\frac{45008}{9}+480 C+1152 C^2-528 \pi ^2\right) \epsilon ^4+
 \left(\frac{58204}{9}+\frac{584 C}{3}+1536 C^2-704 \pi ^2\right) \epsilon ^3 \eta 
- \\ &\left(2256-\frac{464 C}{3}+512 C^2-\frac{752 \pi ^2}{3}\right) \epsilon ^2 \eta ^2+
\left(\frac{520}{3}-\frac{104 C}{3}+32 C^2-\frac{56 \pi ^2}{3}\right) \epsilon  \eta ^3- \\ &
\left(\frac{5942}{9}-\frac{76 C}{3}+192 C^2-76 \pi ^2\right) \epsilon ^2 \xi ^2+
\left(\frac{2902}{9}-\frac{178 C}{3}+74 C^2-\frac{223 \pi ^2}{6}\right) \epsilon  \eta  \xi ^2+\\ &
\left(-\frac{106}{9}+\frac{10 C}{3}-C^2+\frac{13 \pi ^2}{12}\right) \eta ^2 \xi ^2-
\left(\frac{86}{9}-\frac{8 C}{3}+C^2-\frac{13 \pi ^2}{12}\right) \xi ^4
- \\ &
\left(\frac{304}{9}-\frac{22 C}{3}+18 C^2-\frac{9 \pi ^2}{2}\right) \epsilon  \sigma ^3
+\left(-10+\frac{14 C}{3}-3 C^2+\frac{5 \pi ^2}{4}\right) \eta  \sigma ^3\,,
 \end{split}
\end{equation}
\begin{equation}
 \begin{split}
 \kappa_s= \;&-192 \epsilon ^3+192 \epsilon ^2 \eta -32 \epsilon  \eta ^2-24 \epsilon  \xi ^2+2 \eta  \xi ^2+2 \sigma ^3
-  \\ & 
(480+2304 C) \epsilon ^4+\left(\frac{584}{3}+3072 C\right) \epsilon ^3 \eta +\left(\frac{464}{3}-1024 C\right) \epsilon ^2 \eta ^2- \\ &\left(\frac{104}{3}- 64 C\right) \epsilon  \eta ^3+ \left(\frac{76}{3}-384 C\right) \epsilon ^2 \xi ^2-\left(\frac{178}{3}-148 C\right) \epsilon  \eta  \xi ^2+ \\ &
\left(\frac{10}{3}-2 C\right) \eta ^2 \xi ^2+ \left(\frac{8}{3}-2 C\right) \xi ^4-
\left(\frac{22}{3}-36 C\right) \epsilon  \sigma ^3+\left(\frac{14}{3}-6 C\right) \eta  \sigma ^3 -\\ &
\left(\frac{722336}{9}+8448 C+18432 C^2-8448 \pi ^2\right) \epsilon ^5
+\\ &\left(\frac{1175984}{9}+\frac{20464 C}{3}+30720 C^2-14080 \pi ^2\right) \epsilon ^4 \eta
- \\ & \left(\frac{593296}{9}-\frac{4112 C}{3}+15360 C^2-7232 \pi ^2\right) \epsilon ^3 \eta ^2
+ \\ & \left(10840-\frac{3856 C}{3}+2368 C^2-\frac{3568 \pi ^2}{3}\right) \epsilon ^2 \eta ^3
- \\  & \left(\frac{3256}{9}-\frac{272 C}{3}+64 C^2-\frac{112 \pi ^2}{3}\right) \epsilon  \eta ^4
- \\ & \left(\frac{43240}{3}+\frac{56 C}{3}+3840 C^2-1616 \pi ^2\right) \epsilon ^3 \xi ^2
+ \\ & \left(\frac{99580}{9}-1108 C+2724 C^2-1253 \pi ^2\right) \epsilon ^2 \eta  \xi ^2
- \\ & \left(\frac{14470}{9}-\frac{1072 C}{3}+326 C^2-\frac{1057 \pi ^2}{6}\right) \epsilon  \eta ^2 \xi ^2
+ \\ & \left(\frac{118}{9}-4 C+C^2-\frac{13 \pi ^2}{12}\right) \eta ^3 \xi ^2
-  \left(\frac{1202}{3}-80C+82 C^2-\frac{275 \pi ^2}{6}\right) \epsilon  \xi ^4
+ \\ & \left(\frac{394}{9}-\frac{38 C}{3}+4 C^2-\frac{13 \pi ^2}{3}\right) \eta  \xi ^4
+ \left(\frac{8968}{9}-\frac{260 C}{3}+372 C^2-121 \pi ^2\right) \epsilon ^2 \sigma ^3
- \\ & \left(\frac{4858}{9}-136 C+170 C^2-\frac{391 \pi ^2}{6}\right) \epsilon  \eta  \sigma ^3
+ \\ & \left(\frac{302}{9}-\frac{44 C}{3}+7 C^2-\frac{43 \pi ^2}{12}\right) \eta ^2 \sigma ^3
+  \left(\frac{88}{3}-10 C+5 C^2-\frac{41 \pi ^2}{12}\right) \xi ^2 \sigma ^3\,.
  \end{split}
\end{equation}
\vspace{0.5ex}

Using the results from \cite{Lidsey:1995np} and \cite{Gong:2001he} we can
obtain a second order expression for the tensor-to-scalar ratio:
\begin{equation}
r=16 \epsilon -\left(\frac{64}{3}-64 C\right) \epsilon ^2+\left(\frac{32}{3}-32 C\right) \epsilon  \eta\,.
\end{equation}

\end{appendix}

\clearpage

\addcontentsline{toc}{section}{References}
\bibliographystyle{utphys}
\bibliography{inflation_ref_2}

\providecommand{\href}[2]{#2}\begingroup\raggedright\begin{thebibliography}{10}

\bibitem{Ade:2015lrj}
Planck Collaboration, P.~Ade {\em et al.}, ``{Planck 2015. XX. Constraints on
  inflation}'',
\href{http://arxiv.org/abs/1502.02114}{{\tt arXiv:1502.02114 [astro-ph.CO]}}.

\bibitem{Martin:2014vha}
J.~Martin, C.~Ringeval, and V.~Vennin, ``{Encyclop{\ae}dia Inflationaris}'',
  \href{http://dx.doi.org/10.1016/j.dark.2014.01.003}{{\em Phys. Dark Univ.}
  {\bf 5-6} (2014)  75--235},
\href{http://arxiv.org/abs/1303.3787}{{\tt arXiv:1303.3787 [astro-ph.CO]}}.

\bibitem{Okada:2014lxa}
N.~Okada, V.~N. Şenoğuz, and Q.~Shafi, ``{Simple Inflationary Models in Light
  of BICEP2: an Update}'',
\href{http://arxiv.org/abs/1403.6403}{{\tt arXiv:1403.6403 [hep-ph]}}.

\bibitem{Ade:2015tva}
BICEP2, Planck, P.~Ade {\em et al.}, ``{Joint Analysis of BICEP2/Keck Array and
  Planck Data}'', \href{http://dx.doi.org/10.1103/PhysRevLett.114.101301}{{\em
  Phys. Rev. Lett.} {\bf 114} (2015)  101301},
\href{http://arxiv.org/abs/1502.00612}{{\tt arXiv:1502.00612 [astro-ph.CO]}}.

\bibitem{EssingerHileman:2010hh}
T.~Essinger-Hileman, J.~Appel, J.~Beall, H.~Cho, J.~Fowler, {\em et al.},
  ``{The Atacama B-Mode Search: CMB Polarimetry with Transition-Edge-Sensor
  Bolometers}'',
\href{http://arxiv.org/abs/1008.3915}{{\tt arXiv:1008.3915 [astro-ph.IM]}}.

\bibitem{Errard:2010bn}
J.~Errard, ``{The new generation CMB B-mode polarization experiment:
  POLARBEAR}'', in {\em {2010 Rencontres de Moriond proceedings}}.
\newblock 2010.
\newblock
\href{http://arxiv.org/abs/1011.0763}{{\tt arXiv:1011.0763 [astro-ph.IM]}}.
\newblock

\bibitem{Sheehy:2011yf}
C.~Sheehy, P.~Ade, R.~Aikin, M.~Amiri, S.~Benton, {\em et al.}, ``{The Keck
  Array: a pulse tube cooled CMB polarimeter}'',
\href{http://arxiv.org/abs/1104.5516}{{\tt arXiv:1104.5516 [astro-ph.IM]}}.

\bibitem{Fukugita:1986hr}
M.~Fukugita and T.~Yanagida, ``{Baryogenesis Without Grand Unification}'',
\href{http://dx.doi.org/10.1016/0370-2693(86)91126-3}{{\em Phys.Lett.} {\bf
  B174} (1986)  45}.

\bibitem{Buchmuller:2004nz}
W.~Buchmuller, P.~Di~Bari, and M.~Plumacher, ``{Leptogenesis for
  pedestrians}'', \href{http://dx.doi.org/10.1016/j.aop.2004.02.003}{{\em
  Annals Phys.} {\bf 315} (2005)  305--351},
\href{http://arxiv.org/abs/hep-ph/0401240}{{\tt arXiv:hep-ph/0401240
  [hep-ph]}}.

\bibitem{Fayet:1981sq}
P.~Fayet, ``Experimental consequences of supersymmetry'', in {\em Proceedings
  of the XVIth Rencontre de Moriond}, J.~{Tran Thanh Van}, ed., vol.~1,
  pp.~347--367.
\newblock Editions Frontieres,
1981.
\newblock

\bibitem{Pagels:1981ke}
H.~Pagels and J.~R. Primack, ``{Supersymmetry, Cosmology and New TeV
  Physics}'',
\href{http://dx.doi.org/10.1103/PhysRevLett.48.223}{{\em Phys.Rev.Lett.} {\bf
  48} (1982)  223}.

\bibitem{Weinberg:1982zq}
S.~Weinberg, ``{Cosmological Constraints on the Scale of Supersymmetry
  Breaking}'',
\href{http://dx.doi.org/10.1103/PhysRevLett.48.1303}{{\em Phys.Rev.Lett.} {\bf
  48} (1982)  1303}.

\bibitem{Ellis:1984er}
J.~R. Ellis, D.~V. Nanopoulos, and S.~Sarkar, ``{The Cosmology of Decaying
  Gravitinos}'',
\href{http://dx.doi.org/10.1016/0550-3213(85)90306-2}{{\em Nucl.Phys.} {\bf
  B259} (1985)  175}.

\bibitem{Kawasaki:2004yh}
M.~Kawasaki, K.~Kohri, and T.~Moroi, ``{Hadronic decay of late - decaying
  particles and Big-Bang Nucleosynthesis}'',
  \href{http://dx.doi.org/10.1016/j.physletb.2005.08.045}{{\em Phys.Lett.} {\bf
  B625} (2005)  7--12},
\href{http://arxiv.org/abs/astro-ph/0402490}{{\tt arXiv:astro-ph/0402490
  [astro-ph]}}.

\bibitem{Kawasaki:2004qu}
M.~Kawasaki, K.~Kohri, and T.~Moroi, ``{Big-Bang nucleosynthesis and hadronic
  decay of long-lived massive particles}'',
  \href{http://dx.doi.org/10.1103/PhysRevD.71.083502}{{\em Phys.Rev.} {\bf D71}
  (2005)  083502},
\href{http://arxiv.org/abs/astro-ph/0408426}{{\tt arXiv:astro-ph/0408426
  [astro-ph]}}.

\bibitem{Jedamzik:2006xz}
K.~Jedamzik, ``{Big bang nucleosynthesis constraints on hadronically and
  electromagnetically decaying relic neutral particles}'',
  \href{http://dx.doi.org/10.1103/PhysRevD.74.103509}{{\em Phys.Rev.} {\bf D74}
  (2006)  103509},
\href{http://arxiv.org/abs/hep-ph/0604251}{{\tt arXiv:hep-ph/0604251
  [hep-ph]}}.

\bibitem{Moroi:1993mb}
T.~Moroi, H.~Murayama, and M.~Yamaguchi, ``{Cosmological constraints on the
  light stable gravitino}'',
\href{http://dx.doi.org/10.1016/0370-2693(93)91434-O}{{\em Phys.Lett.} {\bf
  B303} (1993)  289--294}.

\bibitem{Pradler:2006hh}
J.~Pradler and F.~D. Steffen, ``{Constraints on the Reheating Temperature in
  Gravitino Dark Matter Scenarios}'',
  \href{http://dx.doi.org/10.1016/j.physletb.2007.02.072}{{\em Phys. Lett.}
  {\bf B648} (2007)  224--235},
\href{http://arxiv.org/abs/hep-ph/0612291}{{\tt arXiv:hep-ph/0612291}}.

\bibitem{Asaka:2000zh}
T.~Asaka, K.~Hamaguchi, and K.~Suzuki, ``{Cosmological gravitino problem in
  gauge mediated supersymmetry breaking models}'',
  \href{http://dx.doi.org/10.1016/S0370-2693(00)00959-X}{{\em Phys. Lett.} {\bf
  B490} (2000)  136--146},
\href{http://arxiv.org/abs/hep-ph/0005136}{{\tt arXiv:hep-ph/0005136}}.

\bibitem{Steffen:2008bt}
F.~D. Steffen, ``{Probing the Reheating Temperature at Colliders and with
  Primordial Nucleosynthesis}'',
  \href{http://dx.doi.org/10.1016/j.physletb.2008.09.036}{{\em Phys. Lett.}
  {\bf B669} (2008)  74--80},
\href{http://arxiv.org/abs/0806.3266}{{\tt arXiv:0806.3266 [hep-ph]}}.

\bibitem{Covi:2010au}
L.~Covi, M.~Olechowski, S.~Pokorski, K.~Turzynski, and J.~D. Wells,
  ``{Supersymmetric mass spectra for gravitino dark matter with a high
  reheating temperature}'',
  \href{http://dx.doi.org/10.1007/JHEP01(2011)033}{{\em JHEP} {\bf 1101} (2011)
   033},
\href{http://arxiv.org/abs/1009.3801}{{\tt arXiv:1009.3801 [hep-ph]}}.

\bibitem{Roszkowski:2012nq}
L.~Roszkowski, S.~Trojanowski, K.~Turzynski, and K.~Jedamzik, ``{Gravitino dark
  matter with constraints from Higgs boson mass and sneutrino decays}'',
  \href{http://dx.doi.org/10.1007/JHEP03(2013)013}{{\em JHEP} {\bf 1303} (2013)
   013},
\href{http://arxiv.org/abs/1212.5587}{{\tt arXiv:1212.5587}}.

\bibitem{Heisig:2013sva}
J.~Heisig, ``{Gravitino LSP and leptogenesis after the first LHC results}'',
  \href{http://dx.doi.org/10.1088/1475-7516/2014/04/023}{{\em JCAP} {\bf 1404}
  (2014)  023},
\href{http://arxiv.org/abs/1310.6352}{{\tt arXiv:1310.6352 [hep-ph]}}.

\bibitem{Lesgourgues:2007gp}
J.~Lesgourgues and W.~Valkenburg, ``{New constraints on the observable inflaton
  potential from WMAP and SDSS}'',
  \href{http://dx.doi.org/10.1103/PhysRevD.75.123519}{{\em Phys.Rev.} {\bf D75}
  (2007)  123519},
\href{http://arxiv.org/abs/astro-ph/0703625}{{\tt arXiv:astro-ph/0703625
  [ASTRO-PH]}}.

\bibitem{Kinney:2002qn}
W.~H. Kinney, ``{Inflation: Flow, fixed points and observables to arbitrary
  order in slow roll}'',
  \href{http://dx.doi.org/10.1103/PhysRevD.66.083508}{{\em Phys.Rev.} {\bf D66}
  (2002)  083508},
\href{http://arxiv.org/abs/astro-ph/0206032}{{\tt arXiv:astro-ph/0206032
  [astro-ph]}}.

\bibitem{Martin:2010kz}
J.~Martin and C.~Ringeval, ``{First CMB Constraints on the Inflationary
  Reheating Temperature}'',
  \href{http://dx.doi.org/10.1103/PhysRevD.82.023511}{{\em Phys.Rev.} {\bf D82}
  (2010)  023511},
\href{http://arxiv.org/abs/1004.5525}{{\tt arXiv:1004.5525 [astro-ph.CO]}}.

\bibitem{Cook:2015vqa}
J.~L. Cook, E.~Dimastrogiovanni, D.~A. Easson, and L.~M. Krauss, ``{Reheating
  predictions in single field inflation}'',
  \href{http://dx.doi.org/10.1088/1475-7516/2015/04/047}{{\em JCAP} {\bf 1504}
  (2015)  047},
\href{http://arxiv.org/abs/1502.04673}{{\tt arXiv:1502.04673 [astro-ph.CO]}}.

\bibitem{Martin:2014nya}
J.~Martin, C.~Ringeval, and V.~Vennin, ``{Observing Inflationary Reheating}'',
  \href{http://dx.doi.org/10.1103/PhysRevLett.114.081303}{{\em Phys.Rev.Lett.}
  {\bf 114} (2015) no.~8, 081303},
\href{http://arxiv.org/abs/1410.7958}{{\tt arXiv:1410.7958 [astro-ph.CO]}}.

\bibitem{Guth:1980zm}
A.~H. Guth, ``{The Inflationary Universe: A Possible Solution to the Horizon
  and Flatness Problems}'',
\href{http://dx.doi.org/10.1103/PhysRevD.23.347}{{\em Phys.Rev.} {\bf D23}
  (1981)  347--356}.

\bibitem{Lucchin:1984yf}
F.~Lucchin and S.~Matarrese, ``{Power Law Inflation}'',
\href{http://dx.doi.org/10.1103/PhysRevD.32.1316}{{\em Phys.Rev.} {\bf D32}
  (1985)  1316}.

\bibitem{Linde:1983gd}
A.~D. Linde, ``{Chaotic Inflation}'',
\href{http://dx.doi.org/10.1016/0370-2693(83)90837-7}{{\em Phys.Lett.} {\bf
  B129} (1983)  177--181}.

\bibitem{Kofman:1996mv}
L.~A. Kofman, ``{The Origin of matter in the universe: Reheating after
  inflation}'',
\href{http://arxiv.org/abs/astro-ph/9605155}{{\tt arXiv:astro-ph/9605155
  [astro-ph]}}.

\bibitem{Dolgov:1982th}
A.~Dolgov and A.~D. Linde, ``{Baryon Asymmetry in Inflationary Universe}'',
\href{http://dx.doi.org/10.1016/0370-2693(82)90292-1}{{\em Phys.Lett.} {\bf
  B116} (1982)  329}.

\bibitem{Albrecht:1982mp}
A.~Albrecht, P.~J. Steinhardt, M.~S. Turner, and F.~Wilczek, ``{Reheating an
  Inflationary Universe}'',
\href{http://dx.doi.org/10.1103/PhysRevLett.48.1437}{{\em Phys.Rev.Lett.} {\bf
  48} (1982)  1437}.

\bibitem{Abbott:1982hn}
L.~Abbott, E.~Farhi, and M.~B. Wise, ``{Particle Production in the New
  Inflationary Cosmology}'',
\href{http://dx.doi.org/10.1016/0370-2693(82)90867-X}{{\em Phys.Lett.} {\bf
  B117} (1982)  29}.

\bibitem{Turner:1983he}
M.~S. Turner, ``{Coherent Scalar Field Oscillations in an Expanding
  Universe}'',
\href{http://dx.doi.org/10.1103/PhysRevD.28.1243}{{\em Phys.Rev.} {\bf D28}
  (1983)  1243}.

\bibitem{Felder:2000hj}
G.~N. Felder, J.~Garcia-Bellido, P.~B. Greene, L.~Kofman, A.~D. Linde, {\em et
  al.}, ``{Dynamics of symmetry breaking and tachyonic preheating}'',
  \href{http://dx.doi.org/10.1103/PhysRevLett.87.011601}{{\em Phys.Rev.Lett.}
  {\bf 87} (2001)  011601},
\href{http://arxiv.org/abs/hep-ph/0012142}{{\tt arXiv:hep-ph/0012142
  [hep-ph]}}.

\bibitem{Liddle:1994dx}
A.~R. Liddle, P.~Parsons, and J.~D. Barrow, ``{Formalizing the slow roll
  approximation in inflation}'',
  \href{http://dx.doi.org/10.1103/PhysRevD.50.7222}{{\em Phys. Rev.} {\bf D50}
  (1994)  7222--7232},
\href{http://arxiv.org/abs/astro-ph/9408015}{{\tt arXiv:astro-ph/9408015
  [astro-ph]}}.

\bibitem{Buchmuller:2013lra}
W.~Buchm�ller, V.~Domcke, K.~Kamada, and K.~Schmitz, ``{The Gravitational
  Wave Spectrum from Cosmological $B-L$ Breaking}'',
  \href{http://dx.doi.org/10.1088/1475-7516/2013/10/003}{{\em JCAP} {\bf 1310}
  (2013)  003},
\href{http://arxiv.org/abs/1305.3392}{{\tt arXiv:1305.3392 [hep-ph]}}.

\bibitem{Endo:2007sz}
M.~Endo, F.~Takahashi, and T.~T. Yanagida, ``{Inflaton Decay in
  Supergravity}'', \href{http://dx.doi.org/10.1103/PhysRevD.76.083509}{{\em
  Phys. Rev.} {\bf D76} (2007)  083509},
\href{http://arxiv.org/abs/0706.0986}{{\tt arXiv:0706.0986 [hep-ph]}}.

\bibitem{Drewes:2013iaa}
M.~Drewes and J.~U. Kang, ``{The Kinematics of Cosmic Reheating}'',
  \href{http://dx.doi.org/10.1016/j.nuclphysb.2013.07.009,
  10.1016/j.nuclphysb.2014.09.008}{{\em Nucl.Phys.} {\bf B875} (2013)
  315--350},
\href{http://arxiv.org/abs/1305.0267}{{\tt arXiv:1305.0267 [hep-ph]}}.

\bibitem{Mazumdar:2013gya}
A.~Mazumdar and B.~Zaldivar, ``{Quantifying the reheating temperature of the
  universe}'', \href{http://dx.doi.org/10.1016/j.nuclphysb.2014.07.001}{{\em
  Nucl.Phys.} {\bf B886} (2014)  312--327},
\href{http://arxiv.org/abs/1310.5143}{{\tt arXiv:1310.5143 [hep-ph]}}.

\bibitem{Freese:1990rb}
K.~Freese, J.~A. Frieman, and A.~V. Olinto, ``{Natural inflation with pseudo -
  Nambu-Goldstone bosons}'',
\href{http://dx.doi.org/10.1103/PhysRevLett.65.3233}{{\em Phys.Rev.Lett.} {\bf
  65} (1990)  3233--3236}.

\bibitem{Adams:1992bn}
F.~C. Adams, J.~R. Bond, K.~Freese, J.~A. Frieman, and A.~V. Olinto, ``{Natural
  inflation: Particle physics models, power law spectra for large scale
  structure, and constraints from COBE}'',
  \href{http://dx.doi.org/10.1103/PhysRevD.47.426}{{\em Phys.Rev.} {\bf D47}
  (1993)  426--455},
\href{http://arxiv.org/abs/hep-ph/9207245}{{\tt arXiv:hep-ph/9207245
  [hep-ph]}}.

\bibitem{Kawasaki:2000yn}
M.~Kawasaki, M.~Yamaguchi, and T.~Yanagida, ``{Natural chaotic inflation in
  supergravity}'', \href{http://dx.doi.org/10.1103/PhysRevLett.85.3572}{{\em
  Phys.Rev.Lett.} {\bf 85} (2000)  3572--3575},
\href{http://arxiv.org/abs/hep-ph/0004243}{{\tt arXiv:hep-ph/0004243
  [hep-ph]}}.

\bibitem{Boubekeur:2005zm}
L.~Boubekeur and D.~Lyth, ``{Hilltop inflation}'',
  \href{http://dx.doi.org/10.1088/1475-7516/2005/07/010}{{\em JCAP} {\bf 0507}
  (2005)  010},
\href{http://arxiv.org/abs/hep-ph/0502047}{{\tt arXiv:hep-ph/0502047
  [hep-ph]}}.

\bibitem{Copeland:1994vg}
E.~J. Copeland, A.~R. Liddle, D.~H. Lyth, E.~D. Stewart, and D.~Wands, ``{False
  vacuum inflation with Einstein gravity}'',
  \href{http://dx.doi.org/10.1103/PhysRevD.49.6410}{{\em Phys.Rev.} {\bf D49}
  (1994)  6410--6433},
\href{http://arxiv.org/abs/astro-ph/9401011}{{\tt arXiv:astro-ph/9401011
  [astro-ph]}}.

\bibitem{Linde:1993cn}
A.~D. Linde, ``{Hybrid inflation}'',
  \href{http://dx.doi.org/10.1103/PhysRevD.49.748}{{\em Phys.Rev.} {\bf D49}
  (1994)  748--754},
\href{http://arxiv.org/abs/astro-ph/9307002}{{\tt arXiv:astro-ph/9307002
  [astro-ph]}}.

\bibitem{Brummer:2014wxa}
F.~Brummer, V.~Domcke, and V.~Sanz, ``{GUT-scale inflation with sizeable tensor
  modes}'', {\em JCAP} {\bf 08} (2014)  066,
\href{http://arxiv.org/abs/1405.4868}{{\tt arXiv:1405.4868 [hep-ph]}}.

\bibitem{Starobinsky:1980te}
A.~A. Starobinsky, ``{A New Type of Isotropic Cosmological Models Without
  Singularity}'',
\href{http://dx.doi.org/10.1016/0370-2693(80)90670-X}{{\em Phys.Lett.} {\bf
  B91} (1980)  99--102}.

\bibitem{Ellis:2013xoa}
J.~Ellis, D.~V. Nanopoulos, and K.~A. Olive, ``{No-Scale Supergravity
  Realization of the Starobinsky Model of Inflation}'',
  \href{http://dx.doi.org/10.1103/PhysRevLett.111.129902,
  10.1103/PhysRevLett.111.111301}{{\em Phys.Rev.Lett.} {\bf 111} (2013) no.~12,
  111301},
\href{http://arxiv.org/abs/1305.1247}{{\tt arXiv:1305.1247 [hep-th]}}.

\bibitem{Ellis:2013nxa}
J.~Ellis, D.~V. Nanopoulos, and K.~A. Olive, ``{Starobinsky-like Inflationary
  Models as Avatars of No-Scale Supergravity}'',
  \href{http://dx.doi.org/10.1088/1475-7516/2013/10/009}{{\em JCAP} {\bf 1310}
  (2013)  009},
\href{http://arxiv.org/abs/1307.3537}{{\tt arXiv:1307.3537}}.

\bibitem{Buchmuller:2013zfa}
W.~Buchmuller, V.~Domcke, and K.~Kamada, ``{The Starobinsky Model from
  Superconformal D-Term Inflation}'',
  \href{http://dx.doi.org/10.1016/j.physletb.2013.08.042}{{\em Phys.Lett.} {\bf
  B726} (2013)  467--470},
\href{http://arxiv.org/abs/1306.3471}{{\tt arXiv:1306.3471 [hep-th]}}.

\bibitem{Farakos:2013cqa}
F.~Farakos, A.~Kehagias, and A.~Riotto, ``{On the Starobinsky Model of
  Inflation from Supergravity}'',
  \href{http://dx.doi.org/10.1016/j.nuclphysb.2013.08.005}{{\em Nucl.Phys.}
  {\bf B876} (2013)  187--200},
\href{http://arxiv.org/abs/1307.1137}{{\tt arXiv:1307.1137}}.

\bibitem{Ferrara:2013wka}
S.~Ferrara, R.~Kallosh, and A.~Van~Proeyen, ``{On the Supersymmetric Completion
  of $R+R^2$ Gravity and Cosmology}'',
  \href{http://dx.doi.org/10.1007/JHEP11(2013)134}{{\em JHEP} {\bf 1311} (2013)
   134},
\href{http://arxiv.org/abs/1309.4052}{{\tt arXiv:1309.4052 [hep-th]}}.

\bibitem{planck_legacy_arxiv}
{Planck Collaboration}, {\em {planck legacy archive}},
  http://www.cosmos.esa.int/web/planck/pla.

\bibitem{Planck:2013jfk}
Planck, P.~Ade {\em et al.}, ``{Planck 2013 results. XXII. Constraints on
  inflation}'', \href{http://dx.doi.org/10.1051/0004-6361/201321569}{{\em
  Astron.Astrophys.} {\bf 571} (2014)  A22},
\href{http://arxiv.org/abs/1303.5082}{{\tt arXiv:1303.5082 [astro-ph.CO]}}.

\bibitem{Antusch:2014saa}
S.~Antusch, F.~Cefala, D.~Nolde, and S.~Orani, ``{False vacuum energy dominated
  inflation with large $r$ and the importance of $\kappa_s$}'',
  \href{http://dx.doi.org/10.1088/1475-7516/2014/10/015}{{\em JCAP} {\bf 1410}
  (2014) no.~10, 015},
\href{http://arxiv.org/abs/1406.1424}{{\tt arXiv:1406.1424 [hep-ph]}}.

\bibitem{Liddle:2003as}
A.~R. Liddle and S.~M. Leach, ``{How long before the end of inflation were
  observable perturbations produced?}'',
  \href{http://dx.doi.org/10.1103/PhysRevD.68.103503}{{\em Phys.Rev.} {\bf D68}
  (2003)  103503},
\href{http://arxiv.org/abs/astro-ph/0305263}{{\tt arXiv:astro-ph/0305263
  [astro-ph]}}.

\bibitem{Planck:2015xua}
Planck, P.~Ade {\em et al.}, ``{Planck 2015 results. XIII. Cosmological
  parameters}'',
\href{http://arxiv.org/abs/1502.01589}{{\tt arXiv:1502.01589 [astro-ph.CO]}}.

\bibitem{Barbieri:1987fn}
R.~Barbieri and G.~Giudice, ``{Upper Bounds on Supersymmetric Particle
  Masses}'',
\href{http://dx.doi.org/10.1016/0550-3213(88)90171-X}{{\em Nucl.Phys.} {\bf
  B306} (1988)  63}.

\bibitem{DeBernardis:1900zz}
F.~De~Bernardis, ``{Constraints on the reheating temperature of the universe
  from WMAP-5 and future perspectives}'',
\href{http://dx.doi.org/10.1016/j.nuclphysbps.2009.07.034}{{\em
  Nucl.Phys.Proc.Suppl.} {\bf 194} (2009)  39--44}.

\bibitem{Ichikawa:2005vw}
K.~Ichikawa, M.~Kawasaki, and F.~Takahashi, ``{The Oscillation effects on
  thermalization of the neutrinos in the Universe with low reheating
  temperature}'', \href{http://dx.doi.org/10.1103/PhysRevD.72.043522}{{\em
  Phys.Rev.} {\bf D72} (2005)  043522},
\href{http://arxiv.org/abs/astro-ph/0505395}{{\tt arXiv:astro-ph/0505395
  [astro-ph]}}.

\bibitem{Crowder:2005nr}
J.~Crowder and N.~J. Cornish, ``{Beyond LISA: Exploring future gravitational
  wave missions}'', \href{http://dx.doi.org/10.1103/PhysRevD.72.083005}{{\em
  Phys.Rev.} {\bf D72} (2005)  083005},
\href{http://arxiv.org/abs/gr-qc/0506015}{{\tt arXiv:gr-qc/0506015 [gr-qc]}}.

\bibitem{Kawamura:2011zz}
S.~Kawamura, M.~Ando, N.~Seto, S.~Sato, T.~Nakamura, {\em et al.}, ``{The
  Japanese space gravitational wave antenna: DECIGO}'',
\href{http://dx.doi.org/10.1088/0264-9381/28/9/094011}{{\em Class.Quant.Grav.}
  {\bf 28} (2011)  094011}.

\bibitem{Nakayama:2008wy}
K.~Nakayama, S.~Saito, Y.~Suwa, and J.~Yokoyama, ``{Probing reheating
  temperature of the universe with gravitational wave background}'',
  \href{http://dx.doi.org/10.1088/1475-7516/2008/06/020}{{\em JCAP} {\bf 0806}
  (2008)  020},
\href{http://arxiv.org/abs/0804.1827}{{\tt arXiv:0804.1827 [astro-ph]}}.

\bibitem{Gong:2001he}
J.-O. Gong and E.~D. Stewart, ``{The Density perturbation power spectrum to
  second order corrections in the slow roll expansion}'', {\em Phys. Lett.}
  {\bf B510} (2001)  1--9,
\href{http://arxiv.org/abs/astro-ph/0101225}{{\tt arXiv:astro-ph/0101225
  [astro-ph]}}.

\bibitem{Lidsey:1995np}
J.~E. Lidsey, A.~R. Liddle, E.~W. Kolb, E.~J. Copeland, T.~Barreiro, and
  M.~Abney, ``{Reconstructing the inflation potential : An overview}'',
  \href{http://dx.doi.org/10.1103/RevModPhys.69.373}{{\em Rev. Mod. Phys.} {\bf
  69} (1997)  373--410},
\href{http://arxiv.org/abs/astro-ph/9508078}{{\tt arXiv:astro-ph/9508078
  [astro-ph]}}.

\end{thebibliography}\endgroup

\end{document}